\documentclass[aps,prx,preprint,onecolumn,citeautoscript,footinbib,
eqsecnum,superscriptaddress]{revtex4-2}
\usepackage{amsmath,amssymb,bm,bbm} 
\usepackage{physics}
\usepackage{graphicx}  
\usepackage{float}
\usepackage[caption = false]{subfig}
\usepackage{dsfont} 
\usepackage{color}
\usepackage{xcolor}
\usepackage[papersize={8.5in,11in}]{geometry}
\usepackage[colorlinks=true]{hyperref}  
\usepackage{ulem}
\usepackage[mathscr]{euscript}
\usepackage[section]{placeins}
\hypersetup{
    unicode=false,          
    pdftoolbar=true,        
    pdfmenubar=true,        
    pdffitwindow=false,     
    pdfstartview={FitH},    
    pdfsubject={},   
    pdfcreator={},   
    pdfproducer={}, 
    pdfkeywords={} {} {}, 
    pdfnewwindow=true,      
    colorlinks=true,       
    linkcolor=magenta, 
    citecolor=blue,        
    filecolor=magenta,      
    urlcolor=blue           
} 

\usepackage{amsfonts}
\usepackage{upgreek}
\usepackage{slashed}
\usepackage{latexsym}

\newcommand{\beq}{\begin{equation}}
\newcommand{\eeq}{\end{equation}}
\def\bea{\begin{eqnarray}}
\def\eea{\end{eqnarray}}

\newcommand{\UU}{\operatorname{U}}
\newcommand{\SU}{\operatorname{SU}}
\newcommand{\SO}{\operatorname{SO}}

\newcommand{\vi}{{\boldsymbol{i}}}
\newcommand{\vj}{{\boldsymbol{j}}}

\definecolor{orange}{rgb}{1,0.5,0}

\newcommand{\tj}[6]{ \begin{pmatrix} #1 & #2 & #3 \\ #4 & #5 & #6 \end{pmatrix}  }

\begin{document}

\title{Anisotropic deconfined criticality in Dirac spin liquids}

\author{Henry Shackleton}
\affiliation{Department of Physics, Harvard University, Cambridge MA 02138, USA}

\author{Subir Sachdev}
\affiliation{Department of Physics, Harvard University, Cambridge MA 02138, USA}
\affiliation{School of Natural Sciences, Institute for Advanced Study, Princeton NJ 08540, USA}

\date{\today}

\begin{abstract}
We analyze a Higgs transition from a $\UU(1)$ Dirac spin liquid to a gapless $\mathbb{Z}_2$ spin liquid. This $\mathbb{Z}_2$ spin liquid 
is of relevance to the spin $S=1/2$ square lattice antiferromagnet, where recent numerical studies have given evidence for such a phase existing in the regime of high frustration between nearest neighbor and next-nearest neighbor antiferromagnetic interactions (the $J_1$-$J_2$ model), appearing in a parameter regime between the vanishing of N\'eel order and the onset of valence bond solid ordering. The proximate Dirac spin liquid is unstable to monopole proliferation on the square lattice, ultimately leading to N\'eel or valence bond solid ordering. As such, we conjecture that this Higgs transition describes the critical theory separating the gapless $\mathbb{Z}_2$ spin liquid of the $J_1$-$J_2$ model from one of the two proximate ordered phases. The transition into the other ordered phase can be described in a unified manner via a transition into an unstable $\SU(2)$ spin liquid, which we have analyzed in prior work. By studying the deconfined critical theory separating the $\UU(1)$ Dirac spin liquid from the gapless $\mathbb{Z}_2$ spin liquid in a $1/N_f$ expansion, with $N_f$ proportional to the number of fermions, we find a stable fixed point with an anisotropic spinon dispersion and a dynamical critical exponent $z \neq 1$. We analyze the consequences of this anisotropic dispersion by calculating the angular profiles of the equal-time N\'eel and valence bond solid correlation functions, and we find them to be distinct. We also note the influence of the anisotropy on the scaling dimension of monopoles. 
\end{abstract}

\maketitle
\newpage
\tableofcontents

\section{Introduction}
\label{sec:intro}
Quantum antiferromagnetism has been a topic of intense study for many decades, and has led to many new insights into quantum many-body phenomenon. In particular, a new class of quantum phases, known as \textit{quantum spin liquids} (QSLs) \cite{Anderson1973, Zhou2017, Savary2016}, are predicted to emerge in certain parameter regimes of antiferromagnets due to the combination of geometric frustration and quantum fluctuations. A generic feature of these QSLs is the existence of fractionalized excitations, which cannot be created individually by local operators.

A particularly well-studied antiferromagnetic model is the $J_1$-$J_2$ model on the square lattice~\cite{IoffeLarkin88,GSH89,Dagotto89,CCL90,RS91,SR91}, which has nearest-neighbor and next-nearest-neighbor antiferromagnetic exchange interactions with coefficients $J_1$ and $J_2$, respectively. It is known that the ground state of the nearest-neighbor square lattice antiferromagnet ($J_2 = 0$) has long-range N\'eel order {\it i.e.\/} global $\SU(2)$ spin rotation symmetry is broken with the spin expectation value $\langle {\bm S}_{\vi} \rangle = \eta_\vi {\bm N}_0$ , where ${\bm S}_\vi$ is the spin operator on site $\vi$, $\eta_\vi = \pm 1$ on the two checkerboard sublattices, and ${\bm N}_0$ is the antiferromagnetic moment. The next-nearest-neighbor antiferromagnetic interactions compete against this N{\'e}el order, and the nature of this model as a function of $J_2 / J_1$, in particular in the regime of high frustration, $J_2 / J_1 \approx 0.5$, remain a key open question.

\begin{figure}
\centering
\includegraphics[width=5in]{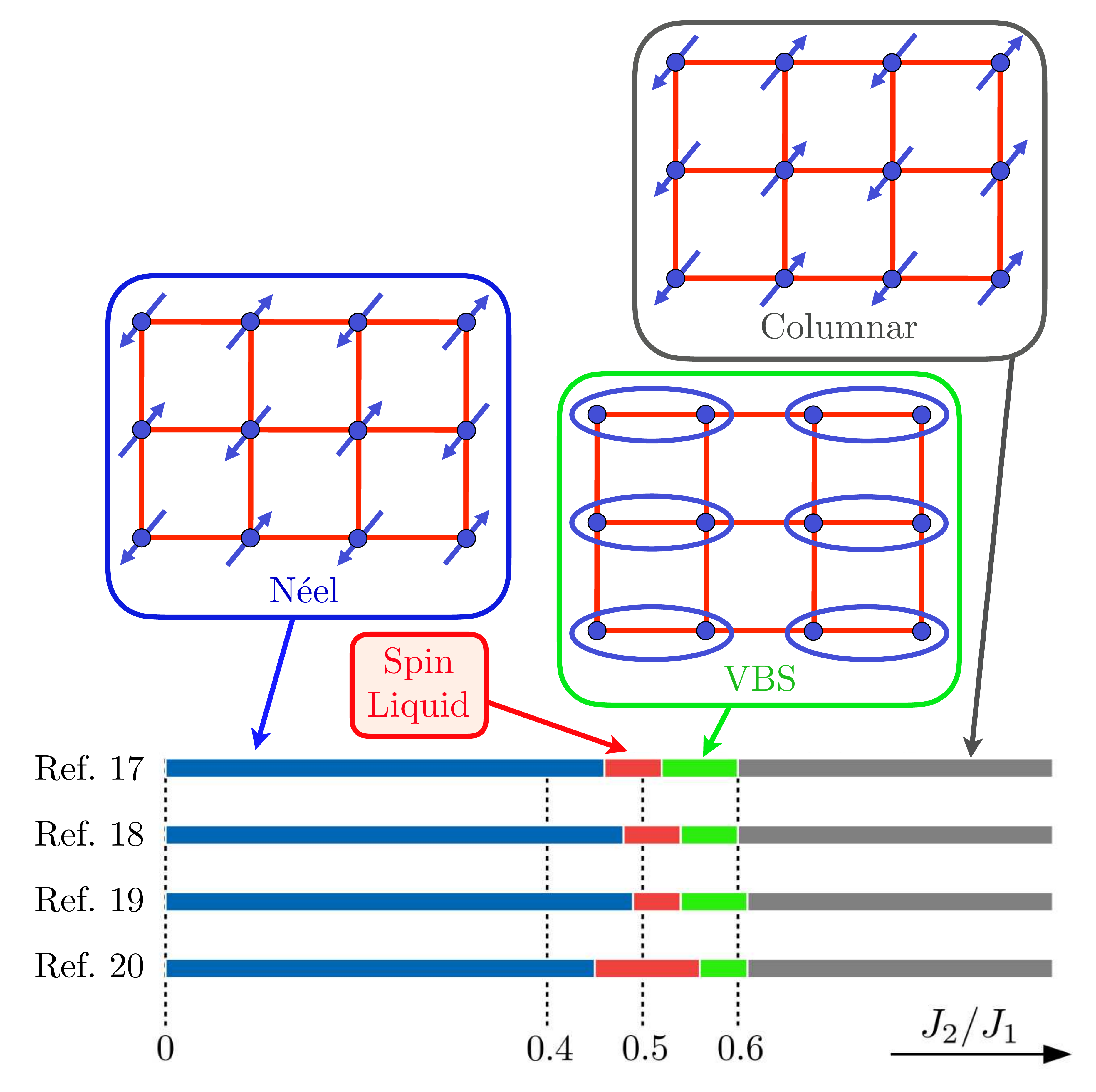}
\caption{Phases of the $S=1/2$ $J_1$-$J_2$ antiferromagnet on the square lattice, from the numerical results of Refs.~\cite{Sandvik18,Becca20,Imada20,Gu20}, all of which agree that the spin liquid is gapless. Each ellipse in the valence bond solid (VBS) represents a singlet pair of electrons. Lower part of figure adapted from Ref.~\cite{BeccaKITP}.}
\label{fig:becca}
\end{figure}

An early proposal \cite{NRSS89,NRSS90,RS91,SR91} was that there was a direct transition from the N\'eel state to a valence bond solid (VBS) (see Fig.~\ref{fig:becca}) which restores spin rotation symmetry but breaks lattice symmetries. This is followed by a first order transition at larger $J_2/J_1$ to a `columnar' state which breaks spin rotation symmetry, and which we do not address in the present work. This led to the development of a theory of `deconfined criticality'~\cite{OMAV04,senthil1,senthil2} which allowed for a direct transition between two symmetry-broken phases, a phenomenon disallowed by conventional Landau-Ginzberg theory. Numerical evidence has since accumulated for the presence of a VBS phase in the $J_1$-$J_2$ model - in particular, it was recently shown that a non-zero antiferromagnetic third-nearest-neighbor interaction $J_3$ stabilizes a clear VBS phase in a large parameter range of $J_2/J_1$~\cite{Liu2021}, and this phase is argued to be stable down to $J_3=0$. The nature of the N\'eel-VBS transition in this model has remained a question of significant debate. However, in the past two years, a consensus appears to have emerged \cite{BeccaKITP} among groups investigating this question by different numerical methods \cite{Sandvik18,Becca20,Imada20,Gu20}, and is summarized in Fig.~\ref{fig:becca}: there is a narrow window with a gapless spin liquid phase between the N\'eel and VBS states. This gapless phase has been identified \cite{Becca01,SenthilIvanov,Becca13,Becca18,Becca20} as a $\mathbb{Z}_2$ spin liquid \cite{RS91,SR91,wen1991,Kitaev1997} with gapless, fermionic, $S=1/2$ spinon excitations with a Dirac-like dispersion \cite{TSMPAF99,WenPSG,SenthilIvanov,SenthilLee05,Kitaev2006}. Although there is less of a consensus over the precise nature of this spin liquid, variational wavefunction studies using a Gutzwiller projection~\cite{Becca18, Becca20} identify the ground state as corresponding to the spin liquid Z2A$zz13$ in Wen's classification \cite{WenPSG}.

In a recent work with A. Thomson~\cite{Shackleton21}, we proposed a unified theory that contains multiple instabilities of the gapless $\mathbb{Z}_2$ spin liquid Z2A$zz13$, which we conjecture correspond to both the neighboring VBS and N\'eel orders shown in Fig.~\ref{fig:becca}. The mechanism for these instabilities consists of considering the $\mathbb{Z}_2$ spin liquid as a condensed phase of a parent $\SU(2)$ gauge theory coupled to Higgs bosons. This $\SU(2)$ theory, identified as the $\pi$-flux spin liquid, also has a proximate phase with $\UU(1)$ gauge symmetry, known as the staggered flux or Dirac spin liquid. Both these phases are conjectured to be unstable on the square lattice and ultimately lead to ordered phases, which gives us a route for explaining the N\'eel and VBS ordered phases predicted to exist alongside the $\mathbb{Z}_2$ spin liquid in the $J_1$-$J_2$ model. In our previous work \cite{Shackleton21}, we studied the transition between the $\SU(2)$ and $\mathbb{Z}_2$ gauge theory using field-theoretic techniques; in this work, we complete our study by an analysis of the $\UU(1)$ to $\mathbb{Z}_2$ transition.

The contents of this work are summarized as follows. In Section~\ref{sec:gaplessz2}, we review the motivation behind our continuum model and explain its connection to the microscopic lattice theory. In Section~\ref{sec:largeN}, we derive a large-$N_f$ effective theory for the $\UU(1)$ to $\mathbb{Z}_2$ transition, with $N_f$ the number of fermion flavors. This allows us to study the critical theory in a systematic ${1}/{N_f}$ expansion. A renormalization-group analysis is performed in Section~\ref{sec:rg}, where we extract critical exponents of the theory to leading order in ${1}/{N_f}$. 

The structure of our critical theory bears resemblence to prior studies of transitions between Dirac spin liquids and a gapped $\mathbb{Z}_2$ gauge theory~\cite{Rufus2018, Dupuis2021}. The difference between this theory and ours is reflected in different forms of Yukawa couplings between the fermions and the Higgs fields. This difference turns out to drastically change the qualitative features of the critical theory. The most notable difference is the lack of Lorentz invariance in our theory. It is known~\cite{Hermele05} that the symmetries of the square lattice permit a velocity anisotropy term in the fermion action. In the absence of additional gapless degrees of freedom, it has been shown~\cite{Franz2002, Vafek2002, Hermele05} that this anisotropy is irrelevant in a $1 / N_f$ expansion. The Yukawa couplings of previously-studied transitions preserve Lorentz invariance. However, we will show that the choice of Yukawa coupling necessary to realize our specific $\mathbb{Z}_2$ spin liquid of interest will lead to Lorentz symmetry breaking and a dynamical critical exponent $z = 1 + 0.225 / N_f + \order{1 / N_f^2}$, where $N_f = 1$ is the physical case.

\section{Summary of prior work and derivation of continuum theory}
\label{sec:gaplessz2}
This paper is a continuation of prior work~\cite{Shackleton21}, which describes multiple possible instabilities of gapless $\mathbb{Z}_2$ spin liquid through a parent $\pi$-flux phase coupled to various Higgs fields. We present a brief summary of this derivation - further details may be found in Ref.~\cite{Shackleton21}.

Our starting point is the fermionic spinon theory of spin liquids, which is derived by re-epressing the spin operators in terms of spinons $f_{\vi \alpha}$, $\alpha = \uparrow\,, \downarrow$ at site $\vi = (i_x\,, i_y)$ of the square lattice using
\begin{equation}
  \begin{aligned}
    \vb{S}_{\vi} &= \frac{1}{2} \sum_{\alpha\,, \beta} f_{\vi \alpha}^\dagger \vb{\sigma}_{\alpha \beta} f_{\vi \beta} \\
  \end{aligned}
\end{equation}
along with the constraints
\begin{equation}
  \begin{aligned}
  f_{\vi \alpha}^\dagger f_{\vi \alpha } = 1 \,, \quad f_{\vi \alpha} f_{\vi \beta} \epsilon_{\alpha \beta } = 0\,.
  \label{eq:halfFillingConstraint}
  \end{aligned}
\end{equation}

Introducing the Nambu spinor,
\begin{equation}
  \begin{aligned}
    \psi_{\vi} &= \begin{pmatrix} f_{\vi \uparrow} \\ f_{\vi \downarrow}^\dagger \end{pmatrix} \,, \\
  \end{aligned}
\end{equation}
and Pauli matrices $\tau^\ell$ which act on spinor indices, we can write a mean-field ansatz for our Hamiltonian
\begin{equation}
  \begin{aligned}
    H &= - \sum_{\vi \vj} \psi_{\vi}^\dagger u_{\vi \vj} \psi_{\vj}\,, \\
  \end{aligned}
\end{equation}
where the hoppings $u_{\vi \vj}$ must be determined self-consistently. The additional degrees of freedom in this representation is reflected by an $\SU(2)_g$ gauge symmetry, under which
\begin{equation}
  \begin{aligned}
    \SU(2)_g : \psi_{\vi} \rightarrow U_{g, \vi} \psi_{\vi} \quad \,, \quad U_{g, \vi} \in \SU(2)
  \end{aligned}
\end{equation}
and a corresponding transformation for $u_{\vi \vj}$. Including gauge fluctuations are necessary to enforce the constraint in Eq.~(\ref{eq:halfFillingConstraint}), which in the Nambu spinor variables becomes $\psi_{\vi}^\dagger \tau^\ell \psi_{\vi} = 0$. Different spin liquids may be described by different mean-field ansatzes $u_{\vi \vj}$, which may also spontaneously break the relevant gauge fluctuations from $\SU(2)$ down to $\UU(1)$ or $\mathbb{Z}_2$.

The particular spin liquid of relevance to the $J_1$-$J_2$ model can be labeled as Z2A$zz$13 following Wen's classification~\cite{WenPSG}. This spin liquid, along with two relevant proximate spin liquids, can be described by the mean-field ansatz,
\begin{equation}
  \begin{aligned}
  u_{\vi,\vi+\hat{x}} &= \chi \, \tau^x - \eta \, \tau^y  \\
u_{\vi,\vi+\hat{y}} &= \chi \, \tau^x + \eta \, \tau^y \\
u_{\vi,\vi+\hat{x}+ \hat{y}} &= - \gamma \, \tau^x  \\
u_{\vi,\vi-\hat{x}+ \hat{y}} &=  \gamma \, \tau^x
\label{eq:mfAnsatz}
  \end{aligned}
\end{equation}
The three relevant spin liquids are:
\begin{itemize}
  \item The $\pi$-flux phase with $\SU(2)$ gauge symmetry corresponds to $\chi = \eta \neq 0$ and $\gamma = 0$.
  \item The staggered flux phase with $\UU(1)$ gauge symmetry corresponds to $\chi\,,\eta \neq 0$, $\gamma = 0$, and $\chi \neq \eta$.
  \item The $\mathbb{Z}_2$ spin liquid Z2A$zz$13 is obtained from the staggered flux phase by turning on a non-zero $\gamma$.
\end{itemize}
The dispersion relation of all three phases hosts 4 Dirac cones at low energy, and hence, all three phases are described by $N = 4$ massless Dirac fermions minimally coupled to the corresponding gauge field. The Dirac cones of the $\pi$-flux phase have an emergent Lorentz invariance, whereas the staggered flux and $\mathbb{Z}_2$ phase have anisotropic dispersion relations on a mean-field level (note that prior studies~\cite{Franz2002, Vafek2002, Hermele05} show an emergent Lorentz invariance of the staggered flux phase upon including gauge fluctuations).

The primary mechanism for our theory of the $J_1$-$J_2$ model is the assumption that both the $\pi$-flux and staggered flux phases on the square lattice are ultimately unstable to ordered phases, either N\'eel or VBS. For the staggered flux phase, the instability arises due to the presence of monopoles, which are allowed by the compactness of the $\UU(1)$ gauge theory. The scaling dimension of these monopoles have been calculated to second-order in a $1/N$ expansion~\cite{Borokhov2002, Pufu2014}, with $N$ the number of fermions, and are relevant for $N=4$. Moreover, there exists a ``trivial'' monopole operator that respects the microscopic symmetries of the square lattice, and hence is allowed by symmetry in an effective Lagrangian~\cite{Alicea2008}. Proliferation of these monopoles is conjectured to lead to ordered phases, including N\'eel and VBS order~\cite{Song1, Song2}. The $\pi$-flux phase, on the other hand, has been conjectured to be a dual description of the DQCP separating N\'eel and VBS order~\cite{Wang17}, and hence is generically unstable to these phases. A unified framework for describing the gapless spin liquid Z2A$zz$13 as well as these two instabilities can hence be obtained from $\text{QCD}_3$ with $N=2$ fermion doublets and an $\SU(2)$ gauge group, and coupling it to Higgs fields whose condensation breaks the gauge group to either $\UU(1)$ or $\mathbb{Z}_2$. The precise manner in which these Higgs fields must couple to the Dirac fermions in order to realize the specific spin liquids of interest can be determined by taking the continuum limit of Eq.~(\ref{eq:mfAnsatz}) and demanding that the condensation of the Higgs fields modifies the Dirac dispersion relation in a manner consistent with the lattice theory. This is the procedure carried out in Ref.~\cite{Shackleton21}. These couplings may also be determined by matching the fractionalization of the square lattice symmetries on the microscopic level with the symmetries of the continuum theory, which is done in Appendix A of Ref.~\cite{Shackleton21}. We refer the reader to Ref.~\cite{Shackleton21} for further details of this calculation; we will simply present the resulting Lagrangian in this work.

This procedure ultimately yields a Lagrangian consisting of four Dirac fermions $\psi$, an $\SU(2)$ gauge field $A_\mu^a$, and three three-component adjoint Higgs fields $\Phi_{1, 2, 3}^a$, $a=x, y, z$. The $\SU(2)$ gauge ($\sigma$) and valley ($\mu$) Pauli matrices both rotate between the four fermion flavors. 
\begin{equation}
  \begin{aligned}
    \mathcal{L} &= \mathcal{L}_{\psi} + \mathcal{L}_{\Phi} + \mathcal{L}_{\Phi \psi}
    \\
    \mathcal{L}_{\psi} &= i \bar{\psi} \gamma^\mu \left( \partial_\mu - i A^a_\mu \sigma^a \right) \psi\,.
    \\
    \mathcal{L}_{\Phi} &= \sum_{i=1}^3 D_\mu \Phi_i^a D^\mu \Phi_i^a + V(\Phi)      
    \\
    \mathcal{L}_{\Phi \psi} &=  \Phi_1^a \, \bar{\psi} \mu^z \gamma^x \sigma^a \psi + \Phi_2^a \, \bar{\psi} \mu^x \gamma^y \sigma^a \psi + \Phi_3^a \bar{\psi} \mu^y \sigma^a (\gamma^x D_y + \gamma^y D_x) \psi
    \\
    V(\Phi) &= s \left( \Phi_1^a \Phi_1^a + \Phi_2^a \Phi_2^a \right) + \widetilde{s} \, \Phi_3^a \Phi_3^a + w \, \epsilon_{abc} \, \Phi_{1}^a \Phi_{2}^b \Phi_{3}^c \nonumber \\
&+ u \left( \Phi_1^a \Phi_1^a + \Phi_2^a \Phi_2^a \right)^2 + \widetilde{u} \left( \Phi_3^a \Phi_3^a \right)^2 + v_1  \left( \Phi_1^a \Phi_2^a \right)^2  +  v_2 \left( \Phi_1^a \Phi_1^a \right)\left( \Phi_2^b \Phi_2^b \right) \nonumber \\
 &+ v_3 \left[ \left( \Phi_1^a \Phi_3^a \right)^2  + \left( \Phi_2^a \Phi_3^a \right)^2 \right]  +  v_4 \left( \Phi_1^a \Phi_1^a + \Phi_2^a \Phi_2^a \right)\left( \Phi_3^b \Phi_3^b \right) + \ldots
    \label{eq:fullLocalLagrangian}
  \end{aligned}
\end{equation}
We have absorbed the coefficients of the Yukawa couplings into the Higgs fields.
The general form of the Higgs potential $V(\Phi)$ is constrained by the microscopic symmetries of the square lattice, and we present only a subset of the possible terms.
The manner in which these microscopic symmetries are embedded in the continuum theory may be derived by starting from the original lattice model, and these transformation properties are given in Table~\ref{tab:symmetryTrans}. The action of the $\SU(2)$ spin rotation symmetry requires a more careful analysis and is described below. All three Higgs fields transform trivially under this $\SU(2)$ symmetry. Our choice of representing the fermionic degrees of freedom in terms of Dirac fermions obfuscates the full $\SU(2)$ spin rotation symmetry, as rotations around the $x$ or $y$ axis involve charge conjugation. However, the $\UU(1)$ subgroup corresponding to rotations around the $z$ axis is simply given by a phase shift in $\psi$, $\psi \rightarrow e^{i\theta} \psi$. The full $\SU(2)$ rotation symmetry may be made explicit by writing the theory in terms of Majorana fermions~\cite{Wang17}, but this will not be necessary for our purposes.
\begin{table}
 \begin{center}
      \begin{tabular}{c||c|c|c|c|c|c}
    & $T_x$ & $T_y$ & $P_x$ & $P_y$ & $\mathcal{T}$ & $R_{\pi/2}$
    \\
    \hline\hline
        $\Phi_1^a$ & $-$ & $+$ & $-$ & $-$ & $-$ & $-\Phi_2^a$ 
    \\
        $\Phi_2^a$ & $+$ & $-$ & $-$ & $-$ & $-$ & $-\Phi_1^a$ 
    \\
        $\Phi_3^a$ &$-$ & $-$& $+$ & $+$ & $+$ & $-$ 
    \\
        $\psi$ & $\mu^x \psi$ & $\mu^z \psi$ & $\gamma^x \mu^z \psi$ & $- \gamma^y \mu^x \psi$ & $\gamma^0 \mu^y \psi$ & $e^{i\pi \gamma^0 / 4} e^{-i \pi \mu^y / 4} \psi$ 
  \end{tabular}
\end{center}
\caption{Listed are the microscopic symmetries of the square lattice and their action on the continuum fields. $T_i$ ($P_i$) indicate translations (reflections) along the $i$'th axis, $\mathcal{T}$ is time-reversal symmetry, and $R_{\pi/2}$ is a $\pi/2$ rotation. We omit the action of $\SU(2)$ spin rotation symmetry as its action on the fermions $\psi$ is non-trivial and described in the main text.}
\label{tab:symmetryTrans}
\end{table}

\begin{figure}[ht]
  \centering
  \includegraphics[width=0.8\textwidth]{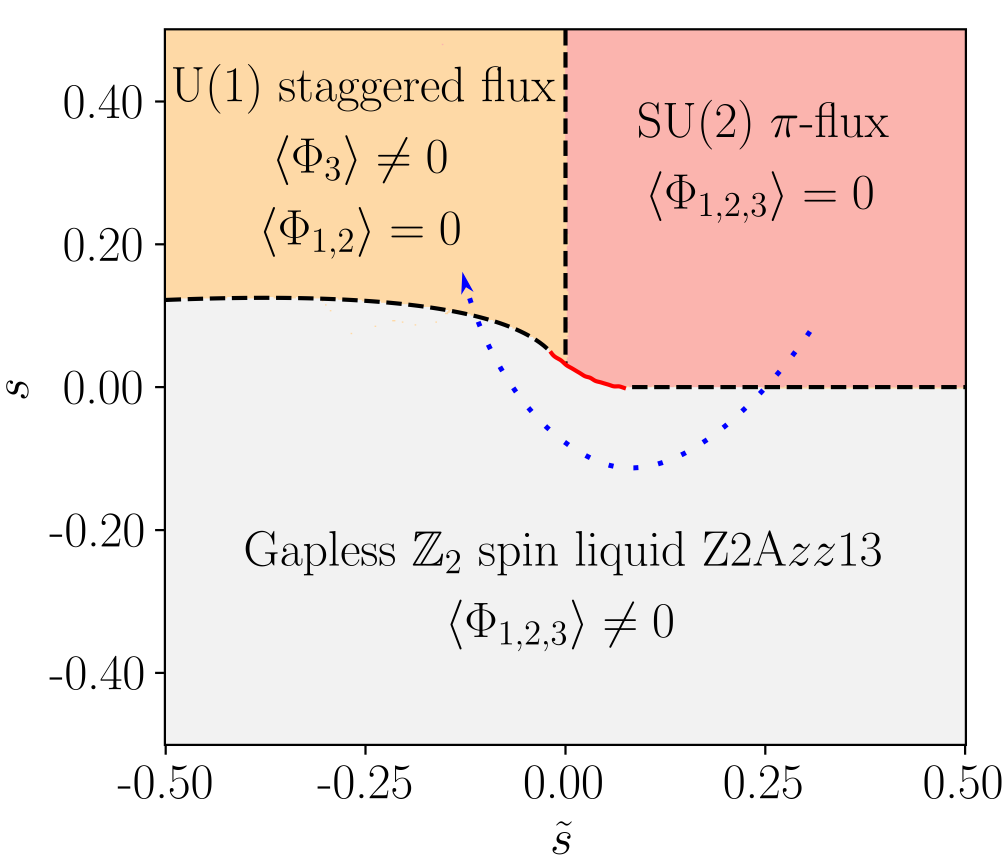}
  \caption{Mean field phase diagram of our low energy theory obtained by minimization of the Higgs potential in Eq.~(\ref{eq:fullLocalLagrangian}). Dashed (solid red) lines indicate second (first) order transitions in mean field theory. We assume the SU(2) $\pi$-flux gauge theory confines to a N\'eel state, the U(1) staggered flux gauge theory confines to a VBS state, except at their deconfined critical boundaries to Wen's stable, gapless $\mathbb{Z}_2$ spin liquid Z2A$zz13$.
    The dotted blue line indicates a possible trajectory of the square lattice antiferromagnet with increasing $J_2/J_1$. }
  \label{fig:mfPhaseDiagram}
\end{figure}
On a mean-field level, this theory admits three phases, illustrated in Fig.~\ref{fig:mfPhaseDiagram}. When all three Higgs fields are uncondensed,  we recover the $\pi$-flux phase with $\SU(2)$ gauge group. When $\Phi_3$ condenses, we obtain the $\UU(1)$ staggered flux phase. The condensation of $\Phi_1$ and $\Phi_2$ yields the gapless spin liquid Z2A$zz$13. The masses of $\Phi_1$ and $\Phi_2$ are fixed to be equal by the microscopic square lattice symmetry, so both condense simultaneously. Furthermore, the symmetry-allowed cubic term $\epsilon_{abc} \Phi_1^a \Phi_2^b \Phi_3^c$ forces $\Phi_3$ to condense along with $\Phi_1$, $\Phi_2$. Our conjectured trajectory of the $J_1$-$J_2$ model as a function of $\frac{J_2}{J_1}$ is shown by the dotted blue line in Fig.~\ref{fig:mfPhaseDiagram}, where transitions into either the staggered flux or $\pi$-flux phases drive the N\'eel or VBS ordering. Our theory may also be compatible with the inclusion of an antiferromagnetic third-nearest-neighbor $J_3$ term, which has been shown numerically~\cite{Liu2021} to compete against the spin liquid phase, eventually leading to a direct N\'eel/VBS transition. This phenomenon can be described in our theory by a deformation of the ${J_2}/{J_1}$ path to a trajectory in the $\pi$-flux phase; this deformation introduces first-order transitions near the tricritical point.

The transition from the $\pi$-flux phase to the $\mathbb{Z}_2$ phase was studied in~\cite{Shackleton21}, and the focus of our work will be the $\UU(1)$ to $\mathbb{Z}_2$ transition.
Before proceeding with our analysis, we briefly summarize the results of our study of the $\SU(2)$ to $\mathbb{Z}_2$ transition. The primary order parameters for this theory are the masses of $\Phi_{1, 2}$, so we neglect fluctuations of $\Phi_3$. This critical theory is studied in a $1/N_f$ expansion, with $4 N_f$ the total number of fermions.  Due to the anisotropic couplings of the Higgs fields, the leading-order effective propagators for $\Phi_{1, 2}$ are divergent along a one-dimensional subspace in momentum space. These lines of zero modes may be thought of as a consequence of an emergent subsystem symmetry. More details on this perspective are presented in Appendix~\ref{app:subsystemSymmetry}. Although these divergences are lifted by higher-order corrections, the leading-order effective action is non-local and more relevant at long distances than the bare Higgs kinetic term. Performing a standard momentum-shell renormalization group study of this leading-order theory is ill-defined and necessitates the inclusion of the irrelevant bare Higgs kinetic terms, which become ``dangerously irrelevant'' due to the singular nature of the leading-order theory. At one-loop order, these features lead to $\log^2$ divergences, rather than the more standard  single-logarithm divergences found in conventional field theories. Re-exponentiating these corrections, we predict that the universal properties of correlation functions at criticality, rather than being power law, are instead $r^{-\alpha} \exp \left( {\beta \ln^2 r} \right)$, with  $\beta= -{6}/{\pi^2}$ for the VBS correlator and $-{12}/{\pi^2}$ for the N\'eel correlator, and $\alpha$ being some \textit{non-universal} coefficient.  

For the remainder of our work, we study the $\UU(1)$ to $\mathbb{Z}_2$ transition, which is driven by the condensation of a charge-2 complex scalar Higgs field. Upon calculating the large-$N_f$ effective action, we perform a renormalization group (RG) analysis to determine the fixed point of our critical theory. The primary observables of interest that we will study are the dynamical critical exponent $z$, which determines the difference between spatial and temporal scaling, and correlations of the N\'eel and VBS order parameters, which are given by fermion bilinears in our continuum theory. 

We find that the presence of the massless Higgs fields strongly modify the critical behavior from the theory of massless QED which describes the pure staggered flux phase without Higgs fields. The staggered flux phase normally possesses an $\SU(4)$ symmetry~\cite{Hermele05}, which relates various physical order parameters including N\'eel and VBS. Moreover, the staggered flux phase has an emergent Lorentz invariance at low energy, as the only velocity anisotropy term allowed by the square lattice symmetries is irrelevant in a $1/N_f$ expansion. The critical Higgs fields explicitly break the $\SU(4)$ symmetry and generate a non-zero velocity anisotropy, which further breaks the $\UU(1)$ spatial rotation symmetry down to the $C_4$ symmetry present in the microscopic model. While correlation functions still have power law decay as in more traditional critical theories, the angular profiles of the correlation functions are modified by the velocity anisotropy; we calculate the tree-level effects of these modifications for the N\'eel and VBS correlation functions.

Recall that in the absence of critical Higgs fields, we postulate that the staggered flux phase is unstable to monopole proliferation. A key assumption in our analysis of the critical theory is that monopoles are rendered irrelevant due to the presence of critical Higgs fields and the theory may be studied using a \textit{non-compact} $\UU(1)$ gauge field. Indeed, such an assumption is similar to the analysis of earlier studies of deconfined criticality between N\'eel and VBS orders~\cite{senthil1, senthil2}. This assumption is elaborated further in  Section~\ref{sec:monopoles}, where we note that in addition to the critical Higgs fields, the presence of a non-zero anisotropy in the fermion dispersion relation can also affect the relevance of monopoles at criticality. In a large-$N_f$ expansion, this has an $\order{N_f}$ effect on the scaling dimension of the monopole, in contrast to the Higgs screening, which is $\order{1}$.

\section{Large \texorpdfstring{$N_f$}{Nf} effective action}
\label{sec:largeN}
Our goal is to study the $\UU(1)$ to $\mathbb{Z}_2$ transition of Eq.~(\ref{eq:fullLocalLagrangian}). Both phases have the Higgs field $\langle \Phi_3^a \rangle \neq 0$, so we fix $\Phi_3^a = \delta_{a z} \Phi$, $\Phi \neq 0$. In this theory, the $\SU(2)$ gauge symmetry is broken to $\UU(1)$, so we only consider $A_\mu \equiv A_\mu^z$. It is important to include the consequences of the cubic Higgs term, whose sign determines the form of the low energy complex Higgs. We choose a gauge where $w \Phi < 0$, and the low energy behavior can be described in terms of a single complex Higgs

\begin{equation}
  \begin{aligned}
    \mathcal{H} &=  \frac{1}{2} \left( \Phi_1^x + \Phi_2^y + i (\Phi_1^y - \Phi_2^x) \right)\,.  \\
    \label{eq:lowEnergyHiggs}
  \end{aligned}
\end{equation}

This field transforms as a charge 2 Higgs field under the unbroken $\UU(1)$ symmetry, as desired. Other linear combinations of $\Phi_{1, 2}$ are massive and can be neglected for the critical theory. As we will clarify later, we must also assume $w < 0$ in order for our transition to be continuous - if $w$ is positive, the aforementioned massive Higgs fields will turn out to have a negative mass at the $\order{1/N_f}$ fixed point, which will lead to a first-order transition.

With these definitions, the Lagrangian that describes the $\UU(1) \rightarrow \mathbb{Z}_2$ transition is
    \begin{equation}
  \begin{aligned}
    \mathcal{L}_{\text{sf}} &= \mathcal{L}_{\psi} + \mathcal{L}_{\mathcal{H}} + \mathcal{L}_{\mathcal{H} \psi}
    \\
    \mathcal{L}_{\psi} &= i\bar{\psi} \gamma^\mu D_\mu \psi +  \Phi \, \bar{\psi} \mu^y \sigma^z \left(\gamma^y D_x + \gamma^x D_y \right) \psi \,.
    \\
    \mathcal{L}_{\mathcal{H}} &= s |\mathcal{H}|^2 + \partial_\mu \mathcal{H}^* \partial^\mu \mathcal{H} + u |\mathcal{H}|^4
    \\
    \mathcal{L}_{\mathcal{H} \psi} &=  \mathcal{H} \bar{\psi} \left(\mu^z \gamma^x + i \mu^x \gamma^y \right) \sigma^-  \psi 
  + \mathcal{H}^\ast \bar{\psi} \left(\mu^z \gamma^x - i \mu^x \gamma^y \right) \sigma^+  \psi\,.
    \label{eq:sfLagrangian}
  \end{aligned}
\end{equation}
with $D_\mu = \partial_\mu - i A_\mu \sigma^z$ and $\sigma^\pm \equiv \sigma^x \pm i \sigma^y$. This Lagrangian may contain higher-order terms, but we omit these as they will turn out to be irrelevant in a $1/N_f$ expansion. The term proportional to $\Phi$ is a modification to the Lorentz-invariant fermion propagator $1 / \slashed{p}$ allowed by the projective symmetry group of the staggered flux phase. In the absence of the critical Higgs field, this velocity anisotropy has a stable fixed point value of $\Phi=0$~\cite{Franz2002, Vafek2002, Hermele05}. In our case, terms of this form are spontaneously generated at one-loop order by the critical Higgs field, and $\Phi$ acquires a non-zero value at the fixed point.

In order to study our critical theory, we proceed in a $1 / N_f $ expansion, with $N_f$ the fermion number. Since our theory only makes sense when the number of fermions $N$ is a multiple of $4$, we define $4 N_f = N$; in other words, we take $N_f = 1$ to correspond to our physical theory. At leading order in ${1}/{N_f}$, our effective  bosonic action takes the form
\begin{equation}
  \begin{aligned}
  \label{eq:effectiveAction}
  \frac{S_{b}}{N_f} &= \int_k \left[s + \Gamma(k)\right]\mathcal{H}^*(-k) \mathcal{H}(k) + \frac{1}{2} \Pi_{\mu\nu}(k) A_\mu(-k) A_\nu(k)  \\
  \end{aligned}
\end{equation}
\begin{figure}[ht]
  \centering
 \includegraphics[width=0.45\textwidth]{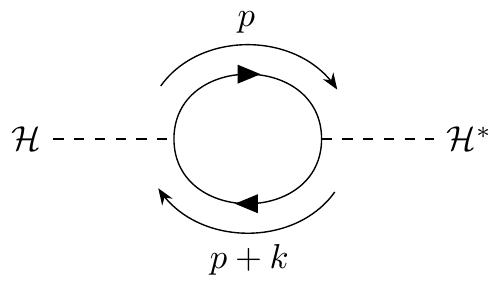}
\includegraphics[width=0.45\textwidth]{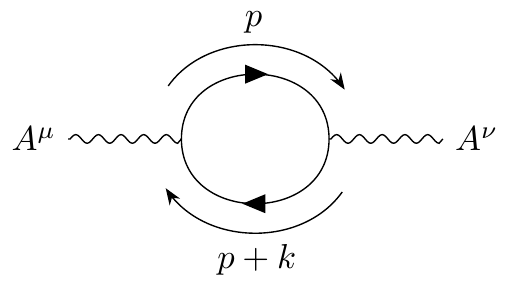}
   \caption{The effective action for the Higgs boson (left) and $\UU(1)$ gauge field (right) are generated by 
   the fermions at leading order in a $1/N_f$ expansion.}
  \label{fig:effectiveActionDiagrams}
\end{figure}
where the inverse propagators $\Gamma$, $\Pi$ are generated by the one-loop fermion diagrams shown in Fig.~\ref{fig:effectiveActionDiagrams}. Note that we have taken the bare Higgs mass to scale with $N_f$, although we will be interested in the critical theory where we tune the Higgs mass to zero. 

To calculate the effective propagators, we need the fermion propagator, which receives corrections to its Lorentz-invariant value of $1/\slashed{p}$ due to a non-zero $\Phi$. This may be treated perturbatively in $\Phi$, but the existence of a stable fixed point turns out to not be viewable at leading order, so we instead proceed with a non-perturbative treatment of $\Phi$. We include further details of this calculation in Appendix~\ref{app:effectiveAction}, and cite the results in the main text. Defining the variables

\begin{equation}
  \begin{aligned}
    k_{x, \pm} \equiv k_x \pm \Phi k_y \,, \quad k_{y, \pm} \equiv k_y \pm \Phi k_x\,, \quad \abs{k_{\pm}} \equiv \sqrt{k_0^2 + k_{x, \pm}^2 + k_{y, \pm}^2}
  \end{aligned}
\end{equation}
the effective inverse Higgs propagator (obtained from the $\Phi$-dependent fermion propagator) is 
\begin{equation}
  \begin{aligned}
    \Gamma(k) &= \frac{1}{16 N_f (1-\Phi^2) } \left[ \frac{ k_+^2 + k_0^2 + 2 k_{x, +} k_{y, +}}{\abs{k_+}} + \frac{ k_-^2 + k_0^2 - 2 k_{x, -} k_{y, -}}{\abs{k_-}}     \right]\,.
    \label{eq:higgsPropIntegrand}
  \end{aligned}
\end{equation}
Likewise, we need the general form of the effective gauge boson propagator. The presence of a non-zero $\Phi$ modifies the gauge coupling, and hence non-Lorentz-invariant corrections arise both from $\Phi$-dependent modifications to the fermion propagator as well as $\order{\Phi}$ vertices. We separate this calculation into three pieces. The first correction comes from using the  $\order{\Phi^0}$ vertices, but with the full fermion propagator. This one-loop term contributes
\begin{equation}
  \begin{aligned}
    \Pi_{\mu\nu}^{(1)}(k) &= \sum_{a = \pm} \frac{k_{a}^2}{8 N_f (1-\Phi^2) \abs{k_{a}}} \left( \delta_{\mu\nu} - \frac{k_{\mu a} k_{\nu a}}{k_{a}^2} \right)\,.
\end{aligned}
\end{equation}
The second correction comes from using one $\order{\Phi}$ vertex, which gives the contribution
\begin{equation}
  \begin{aligned}
    \Pi_{\mu x}^{(2)}(k) &= \Pi^{(2)}_{x \mu}(k) =\sum_{a = \pm} \frac{a \Phi k_{a}^2}{8 N_f (1-\Phi^2) \abs{k_{a}}} \left( \delta_{\mu y} - \frac{k_{\mu a} k_{y a}}{k_{a}^2} \right)  
    \\
    \Pi_{\mu y}^{(2)}(k) &= \Pi_{y \mu}^{(2)}(k) =\sum_{a = \pm}  \frac{a \Phi k_{a}^2}{8 N_f (1-\Phi^2) \abs{k_{a}}} \left( \delta_{\mu x} - \frac{k_{\mu a} k_{x a}}{k_{a}^2} \right)   \,;
  \end{aligned}
\end{equation}
There is also an extra factor of $2$ in $\Pi^{(2)}_{xx, yy}$ due to the two possible vertex orderings. Finally, the third correction comes from using two $\order{\Phi}$ vertices,
\begin{equation}
  \begin{aligned}
    \Pi_{xx}^{(3)}(k) &= \sum_{a=\pm} \frac{\Phi^2 k_{a}^2}{8 N_f (1-\Phi^2) \abs{k_{a}}} \left( 1 - \frac{k_{y a}^2}{k_{a}^2} \right)
    \\
    \Pi_{yy}^{(3)}(k)  &=\sum_{a=\pm} \frac{\Phi^2 k_{a}^2}{8 N_f (1-\Phi^2) \abs{k_{a}}} \left( 1 - \frac{k_{x a}^2}{k_{a}^2} \right)
    \\
    \Pi_{yx}^{(3)}(k) &= \Pi^{(3)}_{xy}(k) = \sum_{a=\pm}-\frac{\Phi^2 k_{x a} k_{y a}}{8 N_f (1-\Phi^2) \abs{k_{a}}} 
  \end{aligned}
\end{equation}
We verify that the combined inverse propagator $\Pi_{\mu\nu}(k) = \sum_{i=1,2,3}\Pi^{(i)}_{\mu\nu}(k)$ annihilates the vector $(k_0, k_x, k_y)$, as required by gauge invariance. Note that $\Pi_{\mu\nu}$ requires a gauge fixing term in order to be invertable. Followin Ref.~\cite{Hermele05}, we add the following non-local gauge fixing term to the Lagrangian
\begin{equation}
  \begin{aligned}
    \frac{1}{4 \xi \abs{k}} A_\mu k^\mu k^\nu A_\nu\,.
  \end{aligned}
\end{equation}
All gauge-invariant observables have been checked to ensure they are independent of the choice of $\xi$.
\section{Renormalization group analysis}
\label{sec:rg}
We perform a renormalization group (RG) analysis of the $\order{1 / N_f}$ effective theory. We are interested in studying the behavior of this theory under the rescaling
\begin{equation}
  \begin{aligned}
    k &= k' e^{-\ell}
    \\
    \omega &= \omega' e^{-z \ell} \\
  \end{aligned}
\end{equation}
We also define a rescaling of the fermion fields
\begin{equation}
  \begin{aligned}
    \psi(k, \omega) &= \psi'(k', \omega') e^{\frac{\ell}{2} (2+2z - \eta_f)}
  \end{aligned}
\end{equation}
The Higgs and gauge fields must also be suitably rescaled, although the anomalous dimensions of these fields will not be needed to calculate our observables of interest. In the absence of a standard boson kinetic term at leading order in ${1}/{N_f}$, we define the scaling of the boson field by performing our RG such that the Yukawa coupling remains fixed under RG.

\subsection{Fermion self-energy}

We first evaluate the $\order{1 / N_f}$ contributions to the fermion self-energy, which come from both gauge and Higgs one-loop diagrams. The self-energy is UV divergent and requires a UV cutoff $\Lambda$. The logarithmic derivative of the fermion self-energy with respect to this cutoff takes the general form
\begin{equation}
  \begin{aligned}
    \Lambda \dv{\Lambda} \Sigma(k) = C_0 k_0 \gamma^0 + C_1 (k_x \gamma^x + k_y \gamma^y) + C_2 \Phi \mu^y \sigma^z (k_x \gamma^y + k_y \gamma^x)
    \label{eq:logDerivativeSelfEnergy}
  \end{aligned}
\end{equation}
for constants $C_{0, 1, 2}$. One must verify that only these terms are generated at one-loop order, which we have done. 

In order to calculate the constants $C_i$, we will use the momentum-shell RG approach outlined in Ref.~\cite{Huh08}. The regularized one-loop contribution to the self-energy schematically takes the form
\begin{equation}
  \begin{aligned}
    \Sigma(k) = \int \frac{\dd[3]{p}}{(2\pi)^3} F(p+k) G(p) \mathcal{K}\left( \frac{p^2}{\Lambda^2} \right) \mathcal{K}\left( \frac{(k+p)^2}{\Lambda^2} \right)
  \end{aligned}
\end{equation}
where $F$ and $G$ are homogeneous functions of the three-momenta, and $\mathcal{K}(y)$ serves as a UV cutoff with the property that $\mathcal{K}(0) = 1$ and $\mathcal{K}(y)$ falls off rapidly for large $y$.  In our calculations, we take $F$ to be the fermion propagator, and $G$ to be the boson propagator (either Higgs or gauge), along with vertex coefficients. The fact that $F$ and $G$ are homogeneous functions allows us to remove the explicit dependence on $\mathcal{K}$ upon taking the logarithmic derivative and integrating by parts. We refer to Appendix~\ref{app:rg} for an explicit derivation of this, and state the result here - the logarithmic derivative of the self-energy takes the form 
\begin{equation}
  \begin{aligned}
    \Lambda \dv{\Lambda} \Sigma(k) &= \frac{k_\lambda}{8\pi^3 N_f}  \int_0^{2\pi} \dd{\phi} \int_0^\pi \sin\theta \dd{\theta} \pdv{F(\hat{p})}{p_\lambda} G(\hat{p})\,.
    \label{eq:rgConvergentIntegrals}
  \end{aligned}
\end{equation}
where $\hat{p} \equiv (\cos\theta, \sin\theta\sin\phi, \sin\theta \cos\phi)$. The resulting integrals in Eq.~(\ref{eq:rgConvergentIntegrals}) are fully convergent and may be evaluated numerically, from which we can extract the coefficients $C_{0, 1, 2}$.

\subsection{Fixed points}
The RG equations for the velocity anisotropy $\Phi$ are
\begin{equation}
  \begin{aligned}
    \dv{\Phi}{\ell} = (C_1 - C_2) \Phi\,.
    \label{eq:rgFlow}
  \end{aligned}
\end{equation}
In the absence of the Higgs field, $\Phi$ has a stable fixed point at $\Phi=0$. The gauge field contribution to this equation has been calculated to leading order in $\Phi$~\cite{Hermele05}, and we verify agreement with this result.

\begin{figure}[ht]
  \centering
  \includegraphics[width=0.5\textwidth]{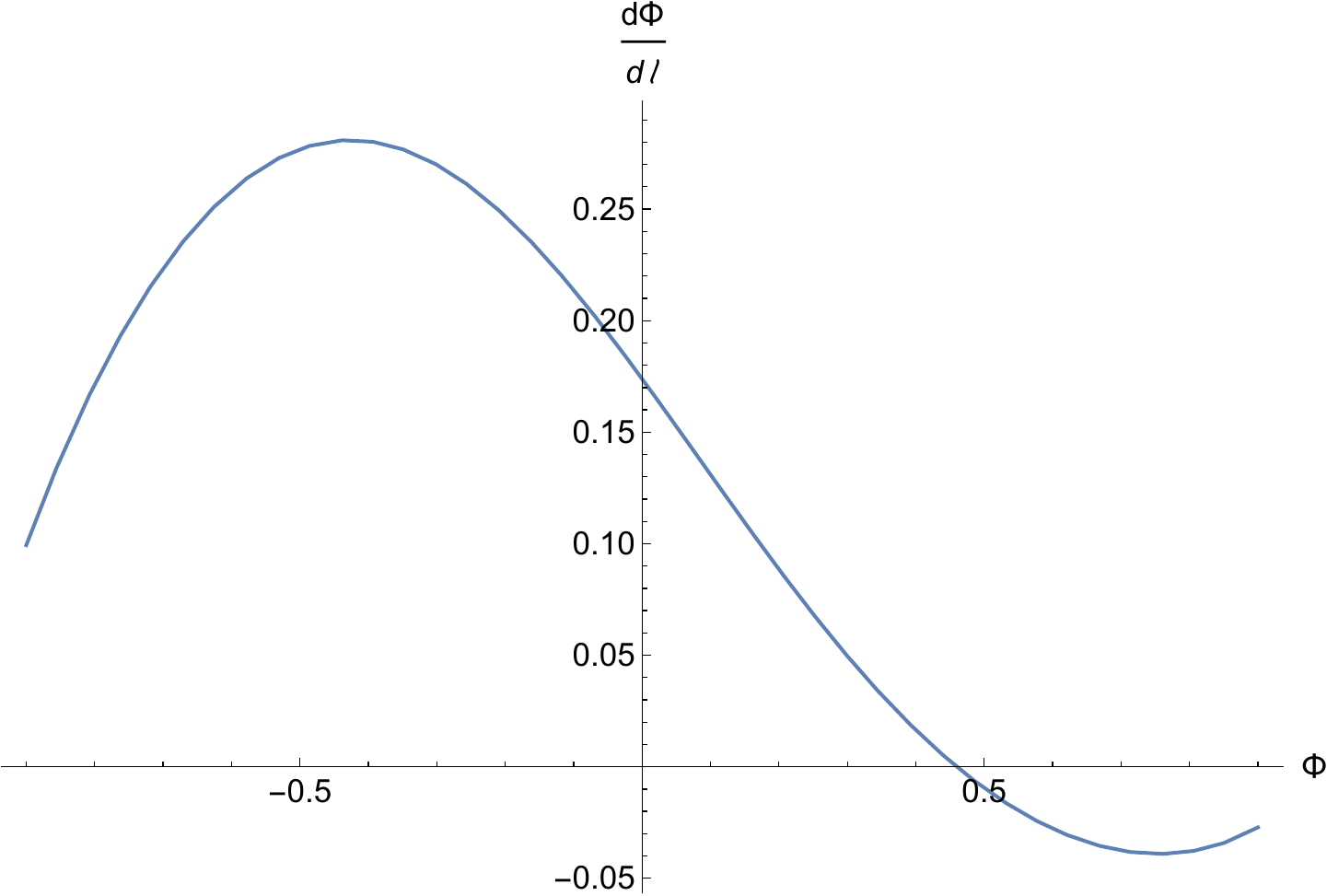}
  \caption{An evaluation of the RG flow of $\Phi$, showing a stable fixed point at $\Phi_c \approx 0.46$.}
  \label{fig:rgFlow}
\end{figure}

The evaluation of Eq.~(\ref{eq:rgFlow}) is plotted in Fig.~\ref{fig:rgFlow}. A stable fixed
point is found at $\Phi_c \approx 0.45765$. At this point, the dynamical critical exponent $z$
is given by
\begin{equation}
  \begin{aligned}
    z = 1 - C_0 + C_1 = 1 + \frac{0.225}{N_f} + \order{1 / N_f^2}\,.
  \end{aligned}
\end{equation}
Recall that when we derived this critical $\UU(1) \rightarrow \mathbb{Z}_2$ theory as a component of a parent $\SU(2)$ theory, we made a gauge choice such that $w \Phi < 0$, where $w$ is the coefficient of the symmetry-allowed cubic term, $w \epsilon_{abc} \Phi_1^a \Phi_2^b \Phi_3^c$. When $\langle \Phi_3^a \rangle = \Phi \delta_{az}$, we can diagonalize this term to yield a mass, $w \Phi (\mathcal{H}^* \mathcal{H} - \mathcal{M}^* \mathcal{M})$, where $\mathcal{H}$ is the combination of $\Phi_{1, 2}^{x, y}$ given in Eq.~\ref{eq:lowEnergyHiggs} and $\mathcal{M}$ is a charge-2 Higgs field of a similar form but with $x \leftrightarrow y$. If we assume $w > 0$, $\Phi < 0$, then $\mathcal{H}$ will become massless first, but the fixed-point value of $\Phi$ gives a negative mass to $\mathcal{M}$, leading to a first-order transition driven by the condensation of $\mathcal{M}$. As a consequence, we must fix our parent $\SU(2)$ theory to have $w < 0$ in order to yield a continuous transition. If we had made a gauge choice such that $w \Phi > 0$, then our theory would have been driven by the condensation of $\mathcal{M}$ rather than $\mathcal{H}$; this still leads to the gapless $\mathbb{Z}_2$ spin liquid Z2Azz13, and all gauge-invariant observables at the critical point remain the same, although the sign of $\Phi_c$ changes.
\subsection{N\'eel and VBS order parameter corrections}
\begin{figure}
    \centering
  \includegraphics[width=1.5in]{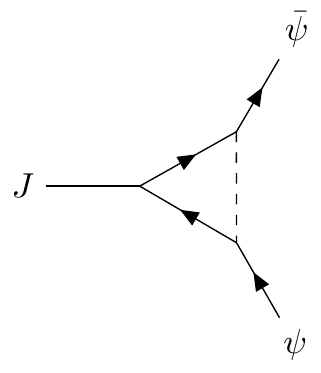}
  \includegraphics[width=1.5in]{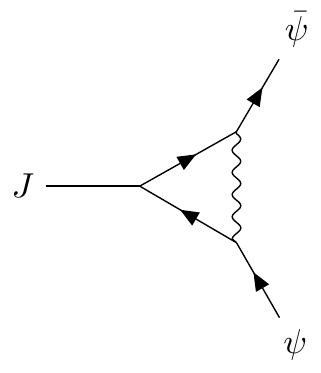}
  \includegraphics[width=2.3in]{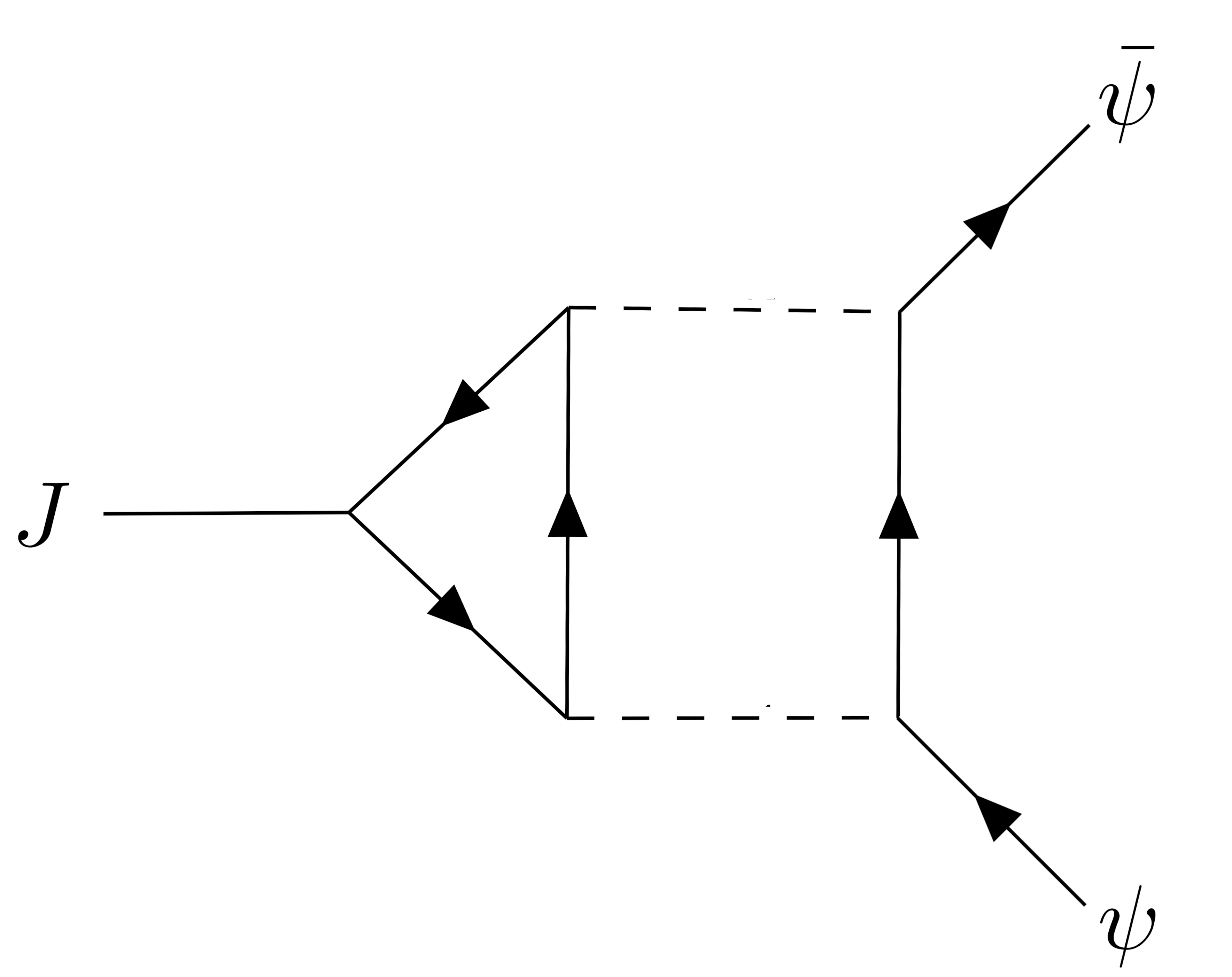}
    \caption{The $\mathcal{O}(N_f^{-1})$ vertex corrections which contribute to the renormalization of the N\'eel and VBS order parameters. 
    The order parameter receives corrections at one-loop order from the Higgs fields (left) and the gauge boson (center). 
  An additional two-loop $\mathcal{O}(N_f^{-1})$ contribution (right) is possible - the diagram shown involves two intermediate
Higgs propagators, but additional diagrams with gauge propagators or mixed gauge/Higgs propagators are possible. These diagrams vanish exactly upon performing the trace over the internal fermion indices.}
    \label{fig:bilinearDiagrams}
\end{figure}

We now calculate the vertex corrections to the N\'eel and VBS order parameters. These order parameters are given by fermion bilinears and can be identified based on the action of the microscopic square lattice symmetries on the fermions. The VBS order parameter is given by the bilinears $\bar{\psi} \mu^{z, x} \psi$. As mentioned previously, our particular representation obfuscates the full $\SU(2)$ action of spin rotation symmetry; however, the $\UU(1)$ subgroup generated by rotations around the $z$-axis is given by the global $\UU(1)$ symmetry $\psi \rightarrow e^{i\theta}\psi$ (recall that this is not the $\UU(1)$ gauge symmetry, which acts as $e^{i \theta \sigma^z}$). As a consequence, we focus on the $z$-component of the N\'eel order parameter, which is given by $\bar{\psi} \mu^y \psi$. The two-point correlation functions of these bilinears are obtained by coupling them to external sources $J_{\text{VBS, N\'eel}}$ and calculating $\order{N_f^{-1}}$ vertex corrections, illustrated in Fig.~\ref{fig:bilinearDiagrams}. Of note are $\order{N_f^{-1}}$ two-loop corrections, the form of which were first found in Ref.~\cite{Hermele05}. These two-loop corrections to the N\'eel and VBS order parameters vanish in the pure staggered flux phase - this follows immediately from taking the trace over the internal fermion loop and noting that $\Tr \mu^i = 0$. We verify that these diagrams remain zero upon the inclusion of both Higgs fields and velocity anisotropy, although this identity is less readily apparent. 

\subsubsection{One-loop vertex corrections}
We first outline the procedure from Ref.~\cite{Huh08} for calculating the logarithmic corrections to the vertex functions.  At zero external momenta, our one-loop vertex corrections schematically take the form
\begin{equation}
  \begin{aligned}
   \Xi_i  = \frac{1}{N_f} \int \frac{\dd[3]{p}}{(2\pi)^3} H_i(p) \mathcal{K}^3\left( \frac{p^2}{\Lambda^2} \right)
  \end{aligned}
\end{equation}
where $H_i(p)$, $i = x, y, z$, is a homogeneous function of $p$ and illustrated in Fig.~\ref{fig:bilinearDiagrams}. The index $i$ indicates whether the $J$ vertex includes a factor of $\mu^{x, z}$ (VBS order parameter) or $\mu^y$ (N\'eel order parameter). Once again, we can take the logarithmic derivative and remove the explicit cutoff dependence, leading to the equation
\begin{equation}
  \begin{aligned}
    \Lambda \dv{\Lambda} \Xi_i = \frac{1}{8\pi^3 N_f}  \int_0^{2\pi} \dd{\phi} \int_0^\pi \sin\theta \dd{\theta} H_i(\hat{p}) \equiv B_i \mu^i\,.
  \end{aligned}
\end{equation}
The $B_i$'s are not gauge-invariant by themselves, and must be combined with the self-energy to get a gauge-invariant quantity, which at the fixed-point value $\Phi_c$ gives
\begin{equation}
  \begin{aligned}
    \eta_{\text{VBS}} &= B_{x, z} + C_0  \approx 0.06468 N_f^{-1} + \order{N_f^{-2}}\,,
    \\
    \eta_{\text{N\'eel}} &= B_{y} + C_0  \approx -0.01634 N_f^{-1} + \order{N_f^{-2}}\,.
  \end{aligned}
\end{equation}

The N\'eel and VBS correlators in momentum space have the scaling form

\begin{equation}
    \begin{aligned}
      G_{\text{N\'eel}}(k, \omega) &=  G_{\text{N\'eel}}(a k, a^z \omega) a^{2 \eta_{\text{N\'eel}} - 1}\,,
      \\
       G_{\text{VBS}}(k, \omega) &=  G_{\text{VBS}}(a k, a^z \omega) a^{2 \eta_{\text{VBS}} - 1}\,.
    \end{aligned}
\end{equation}

Making a Fourier transform to real space, the equal-time N\'eel and VBS correlators have the power law decay

\begin{equation}
    \begin{aligned}
      G_{\text{N\'eel}}(r) &\sim \frac{1}{r^{3 + z - 2 \eta_{\text{N\'eel}}}}
      \\
      G_{\text{VBS}}(r) &\sim \frac{1}{r^{3 + z - 2 \eta_{\text{VBS}}}}\
    \end{aligned}
\end{equation}
Note that both the anomalous dimensions for the N\'eel and VBS correlators are quite small. This is a rather surprising result and does not seem to be due to any particular small parameter. The magnitude of these anomalous dimensions do not decrease upon increasing the numerical precision of our integration, so we believe them to be small but not identically zero. We find that the gauge fluctuations generally enhance N\'eel and VBS correlations, whereas Higgs fluctuations suppress them - the combined result is the stated anomalous dimensions. As such, we cannot make a strong statement regarding which ordering the unstable $\UU(1)$ phase will prefer, as neither the N\'eel nor VBS order parameter show exceptionally enhanced correlations. Higher-order corrections may show a clearer preference to either N\'eel or VBS ordering.

\subsection{Tree-level effect of velocity anisotropy on correlation functions}
As we have emphasized, one of the key features of this critical theory is that the emergent Lorentz invariance of the staggered flux phase is broken by the presence of critical Higgs fields, leading to a non-zero value of the symmetry-allowed velocity anisotropy term. This anisotropy term also has the effect of breaking the emergent $\SU(4)$ flavor symmetry. We refer to Ref.~\cite{Hermele05} for a more extensive study of the intertwining physical order parameters of the $\SU(4)$ theory and which relations hold in the presence of the velocity anisotropy - for our purposes, we note that the emergent $\SO(5) \subset \SU(4)$ symmetry that relates the N\'eel and VBS order parameters is broken down to the microscopic $\SO(3) \times C_4$. At tree-level, the scaling dimensions of the two order parameters are still the same, but the angular profile of their correlation functions are modified due to the velocity anisotropy. This lack of an emergent $\SO(2)$ spatial rotation symmetry in the N\'eel and VBS correlation functions may be useful as a numerical probe of the critical behavior, so we study the angular profile in more detail.

We analytically compute the spatial profile of the N\'eel order parameter at tree level.  This calculation turns out to be feasible non-perturbatively in the velocity anisotropy $\Phi$. The VBS correlator is more difficult to study non-perturbatively in the velocity anisotropy, and we will later compute corrections to leading order in $\Phi$.

\begin{figure}[ht]
  \centering
  \includegraphics[width=0.5\textwidth]{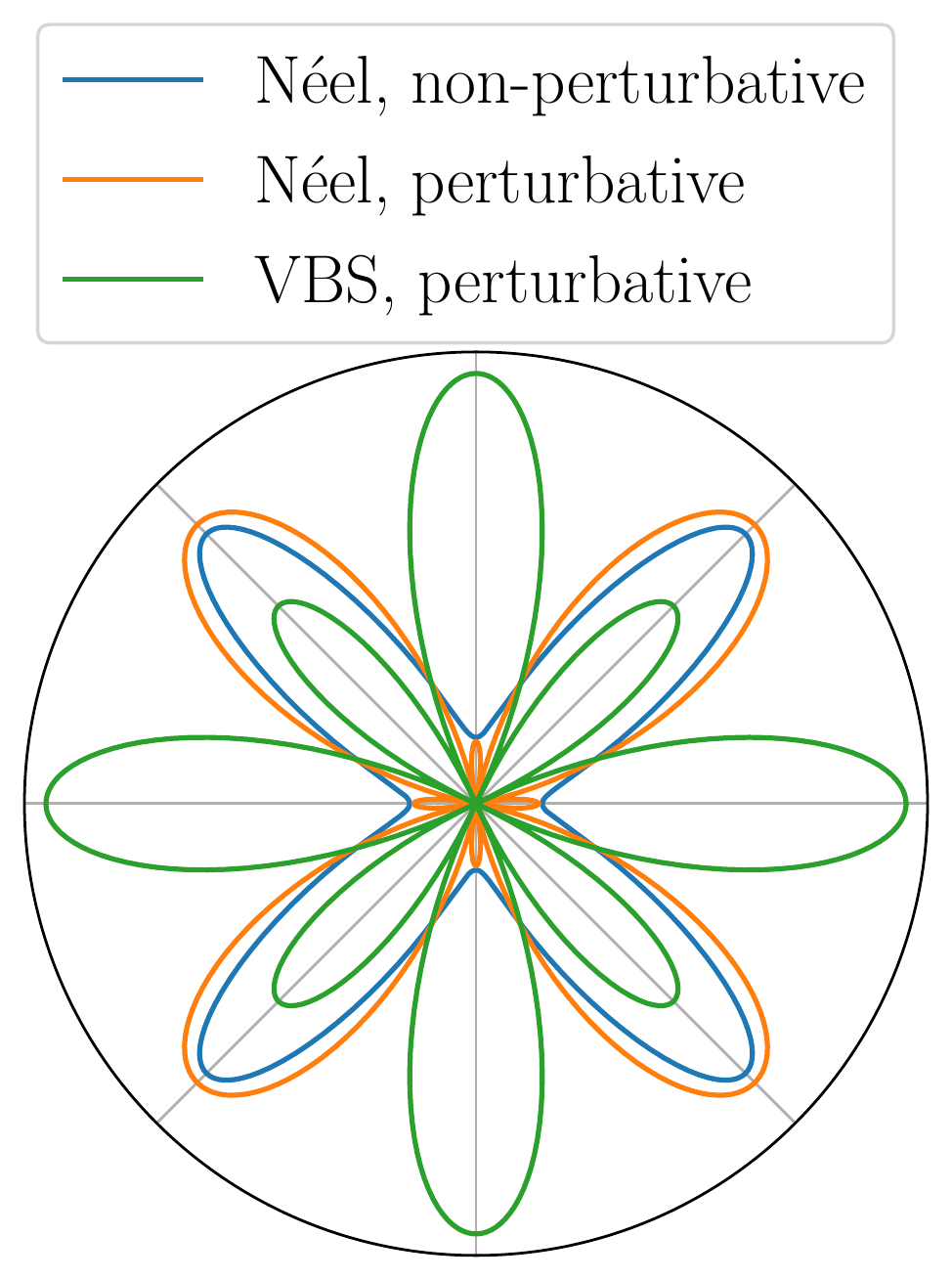}
  \caption{Plotted are the angular profiles of the equal-time N\'eel and VBS correlation functions in real space, at the fixed point value of velocity anisotropy $\Phi_c$. The N\'eel order parameter shows enhanced correlations along the diagonals, whereas the VBS correlations are more enhanced along the $x$ and $y$ directions. Note that we only plot the absolute value of the correlation function, and the signs of the perturbative N\'eel and VBS correlators flip when moving from the $x$ and $y$ axes to the diagonals. As this feature is not present in the non-perturbative N\'eel correlator, it is possible that this feature similarly vanishes at higher orders for the VBS correlator. }
  \label{fig:spatialDependence}
\end{figure}

The two-point function in momentum space is given by, with $Q(p)$ the fermion propagator,
\begin{equation}
  \begin{aligned}
    G_{\text{N\'eel}}(k) &= - \int \frac{\dd[3]{p}}{(2\pi)^3} \Tr\left[ Q(p) \mu^y Q(p + k) \mu^y \right]
    \\
    &= - \frac{2}{1-\Phi^{2}} \sum_{a=\pm}\int \frac{\dd[3]{p}}{(2\pi)^3} \frac{p_0 (p_0 + k_0) + a p_x (p_x + k_{x, a}) + a p_y (p_y + k_{y, a})}{p^2 (p + k_{\pm})^2}
    \\
    &= \frac{1}{8(1-\Phi^2)}\left( \abs{k_+} + \abs{k_-}  \right)\,.
  \end{aligned}
\end{equation}
As before, we define $k_{\pm} = (k_0, k_x \pm \Phi k_y, k_y \pm \Phi k_x)$. The Fourier transform
\begin{equation}
  \begin{aligned}
    \int \frac{\dd[3]{k}}{(2\pi)^3} e^{i k \cdot r} \abs{k_\pm}
  \end{aligned}
\end{equation}
can be performed by a change of variables to give
\begin{equation}
  \begin{aligned}
     G_{\text{N\'eel}}(r) &\sim \sum_{a=\pm} f_{a}(\theta, \phi)^{-2} \frac{1}{r^4}
  \end{aligned}
\end{equation}
where we change to spherical coordinates, $t = r \cos\theta$, $x = r \sin\theta\sin\phi$, $y = r \sin\theta \cos\phi$, and
\begin{equation}
  \begin{aligned}
    f_{\pm}(\theta, \phi) &=  1 + \frac{\sin^2\theta\left( \pm  2 \Phi \sin 2 \phi  + 3 \Phi^2  + \Phi^4 \right)}{(1 - \Phi^2)^2}\,.
  \end{aligned}
\end{equation}
Therefore,
\begin{equation}
  \begin{aligned}
    G_{\text{N\'eel}}(r) \sim g(\theta, \phi) \frac{1}{r^4}
  \end{aligned}
\end{equation}
where $g(\theta, \phi) = {1}/{f_+(\theta, \phi)^2} + {1}/{f_-(\theta, \phi)^2}$ is plotted in Fig.~\ref{fig:spatialDependence}. We note the enhanced correlations of the N\'eel order parameter along the diagonals, which holds true for generic
values of $\Phi$. 

An analogous calculation of the VBS order parameter is less analytically tractable, as the one-loop integral cannot be made isotropic by a coordinate transformation. As such, we resort to a perturbative study of the velocity anisotropy. This gives
\begin{equation}
  \begin{aligned}
    G_{\text{VBS}}(k) \sim \abs{k} - \Phi^2 \left[ 2 \abs{k} + \frac{k_x^2 k_y^2 - k_0^2 k^2}{\abs{k}^3} \right] + \order{\Phi^4}
  \end{aligned}
\end{equation}
in momentum space, or 
\begin{equation}
  \begin{aligned}
    G_{\text{VBS}}(r) \sim \frac{1}{r^4} \left[ 8 + \Phi^2 (\cos 2\theta (40 + 12 \cos 4 \phi) + \cos 4 \theta (6 - 3 \cos 4 \phi) - 18 \cos 4 \phi - 14) \right] 
  \end{aligned}
\end{equation}
in real space. The equal-time VBS correlation function is plotted in Fig.~\ref{fig:spatialDependence}, showing enhanced correlations along the cardinal directions. Note that the correlation function changes sign on the diagonals - this is an unusual feature, and would seemingly indicate lines in real space where the VBS correlator vanishes. This feature is also present in the $\order{\Phi^2}$ corrections to the N\'eel correlator but ultimately vanishes in the non-perturbative result, so this result may only be an artifact of the perturbative expansion. Further details on this calculation can be found in Appendix~\ref{app:ft}. 
\section{Monopoles}
\label{sec:monopoles}
On the square lattice, there exists a monopole operator in the staggered flux phase - the \textit{trivial monopole} - that is invariant under all square lattice symmetries, and hence is an allowed perturbation. To leading order, the scaling dimension of the monopole operator  scales with the number of fermions and becomes irrelevant for $N_f \geq 3$. Hence, the staggered flux phase by itself is unstable to monopole proliferation - this is the mechanism which we claim gives rise to ordered phases in the staggered flux phase, as condensation of the trivial monopole is conjectured to lead to a fermion chiral mass generation corresponding to either N\'eel or VBS order ~\cite{Song1, Song2}. Our calculations of the critical theory rely on the assumption that the presence of massless scalar fields screens monopoles and renders them irrelevant at the critical point. Here, we draw attention to an additional contribution to the monopole scaling dimension, which is the non-zero anisotropy in the spinon dispersion. Prior calculations of monopole scaling dimensions in $\text{QED}_3$ assume a Lorentz-invariant action for the fermions, which is natural in pure $\text{QED}_3$ given that velocity anisotropy terms are irrelevant in a ${1}/{N_f}$ expansion. However, as we have shown, the presence of critical Higgs fields can give rise to a non-zero velocity anisotropy at the critical point. An important question is whether this anisotropy increases or decreases the monopole scaling dimension. In contrast with the direct modification arising from the critical fields, which is $\order{1}$, the effect of the anisotropy on the monopole scaling dimension is $\order{N_f}$. Such a modification, if calculated perturbatively in $\Phi$, arises at $\order{\Phi^2}$ - this is still an appreciable shift given the relatively large anisotropy $\Phi_c \approx 0.46$. Previous works have studied the effects of a spin Hall mass on the monopole scaling dimension~\cite{Dupuis2019,Dupuis2021b}, although this perturbation is more tractable as the spin Hall mass is diagonal in the basis of spinor monopole harmonics. In Appendix~\ref{app:monopole}, we outline the structure of a perturbative calculation for calculating the $\order{\Phi^2}$ corrections to the monopole scaling dimension. An important observation which makes this calculation tractable is that, while the saddle-point monopole gauge configuration will not take the form of the rotationally-invariant Dirac monopole, corrections to the scaling dimension arising from this difference only arise at higher orders in $\Phi$; hence, to lowest non-trivial order, one can assume a Dirac monopole background. This calculation ultimately yields a divergent summation of terms involving Wigner $3$-$j$ symbols; we leave for future work further study of how to properly regularize this calculation. 

Additionally, we briefly comment on the relation between this velocity anisotropy and the monopole quantum numbers. Prior studies on the effects of a spin Hall mass~\cite{Dupuis2019, Dupuis2021b} have found that the presence of such a term induces a spin polarization on the monopoles. Each fermion flavor has a zero mode in the presence of a monopole, and half of these zero modes must be filled in order to maintain gauge neutrality of the monopole. The presence of a spin Hall mass polarizes these zero modes, which in turn causes a splitting in the scaling dimension of the monopoles, with the most-relevant monopole being spin polarized. One may wonder whether a similar valley polarization can arise due to our velocity anisotropy term due to the presence of a $\mu^y$ in the anisotropy; however, we check that the first-order energy splitting of the fermion zero modes due to the velocity anisotropy vanishes. Higher-order corrections including corrections from Higgs and gauge fluctuations will in general break the six-fold degeneracy of monopole scaling dimensions; in particular, the monopoles with N\'eel and VBS quantum numbers will have different scaling dimensions, which may cause a preference towards a particular type of symmetry-breaking in the staggered flux phase. Further study of the spectrum of monopoles at this critical point may be useful for determining the IR fate of the proximate staggered flux phase - our observation is that this behavior is more complicated than a simple valley polarization of the monopoles.

\section{Conclusions and future directions}

We have presented a large-$N_f$ analysis of a deconfined critical theory separating a gapless $\mathbb{Z}_2$ spin liquid Z2Azz13 from the $\UU(1)$ staggered flux phase, the latter of which we assume to be unstable to either N\'eel or VBS ordering on the square lattice. This completes the large-$N$ study of the phase diagram shown in Fig.~\ref{fig:mfPhaseDiagram}, where the gapless $\mathbb{Z}_2$ spin liquid may emerge as a Higgsed phase from either a $\UU(1)$ or $\SU(2)$ gauge theory. Both of these parent gauge theories are conjectured to be unstable on the square lattice, and hence we propose the trajectory through the phase diagram as shown in Fig.~\ref{fig:mfPhaseDiagram} as a description of the $J_1$-$J_2$ square lattice antiferromagnet, where numerical studies suggest a gapless $\mathbb{Z}_2$ spin liquid emerging between N\'eel and VBS phases. 

Our calculations yield several predictions which may be investigated by future numerical studies. One of the most striking features of both the $\SU(2) \rightarrow \mathbb{Z}_2$ and $\UU(1) \rightarrow \mathbb{Z}_2$ transitions is the lack of Lorentz invariance, spatial rotation invariance (aside from the discrete $C_4$ rotational symmetry), and in the case of the $\SU(2) \rightarrow \mathbb{Z}_2$ transition, even a lack of traditional scale invariance. The lack of scale invariance takes the form of correlation functions decaying as $e^{-\ln^2(r)}$ rather than power law, the difference of which is difficult to detect for small system sizes. As such, it may be more promising to search for lack of Lorentz invariance ($z \neq 1$), or lack of a full $\SO(2)$ spatial rotation invariance of correlation functions. We draw attention to the angular profiles of the N\'eel and VBS correlation functions shown in Fig.~\ref{fig:spatialDependence}, which come from a mean-field description of the staggered flux state with the inclusion of a symmetry-allowed velocity anisotropy and predict enhanced N\'eel correlations along the diagonals, and enhanced VBS correlations along the cardinal directions.

\section{Acknowledgements}
We are grateful to Zheng-Cheng Gu, Patrick Ledwith, and Alex Thomson for insightful discussions. This research was supported by the National Science Foundation under Grant No. DMR-2002850. This work was also supported by the Simons Collaboration on Ultra- Quantum Matter, which is a grant from the Simons Foundation (651440, S.S.).

\appendix

\section{Alternative proximate \texorpdfstring{$\mathbb{Z}_2$}{Z2} spin liquids}
\label{app:altSpinLiquids}
The phase diagram of the theory we have defined contains three phases on a mean-field level - the $\SU(2)$ spin liquid of the $\pi$-flux phase, the $\UU(1)$ staggered flux phase, and the gapless $\mathbb{Z}_2$ spin liquid whose projective symmetry group labels it as Z2A$zz$13 according to Wen's classification~\cite{WenPSG}. Although these spin liquids are the ones we believe to be of relevance to the $J_1$-$J_2$ model, additional $\mathbb{Z}_2$ spin liquids are accessible in this general framework by a modification of the Higgs couplings~\cite{Thomson17}. Although there are many $\mathbb{Z}_2$ spin liquids accessible starting from the $\pi$-flux phase and we will not attempt to study all of them, we will identify the continuum theories associated with the eight $\mathbb{Z}_2$ spin liquids proximate to \textit{both} the $\pi$-flux and staggered flux phase.

\subsection{Symmetry fractionalization in the continuum staggered flux theory}
The $\SU(2)$ gauge symmetry introduced in the fermionic spinon theory of spin liquids means that the microscopic symmetries of the square lattice need no longer be realized explicitely, but may be realized \textit{projectively}, i.e., up to an overall $\SU(2)$ gauge transformation. This concept holds true in our continuum theory, and we will use this to identify possible $\mathbb{Z}_2$ spin liquids based off of how the microscopic symmetries are realized. To see this, consider the field theory describing the transition from the $\pi$-flux phase to the staggered flux phase, which consists of four Dirac fermions and a three-component adjoint Higgs field $\Phi_3$, both minimally coupled to an $\SU(2)$ gauge field, and an additional Yukawa coupling
\begin{equation}
  \begin{aligned}
    \Phi_3^a  \overline{\psi} \sigma^a \mu^y (\gamma^x i \partial_y + \gamma^y i \partial_x) \psi \equiv \Phi_3^a  \overline{\psi} \sigma^a M \psi
  \end{aligned}
\end{equation}
When $\Phi_3$ is uncondensed, it can be integrated out, and the Yukawa couplings generate terms irrelevant at long distances. When $\Phi_3$ is condensed - for concreteness, $\langle \Phi_3^a \rangle = \Phi \delta_{a z}$ - we replace $\Phi_3^a$ by its expectation value, and the fermion bilinear $ \overline{\psi} \sigma^z \mu^y (\gamma^x i \partial_y + \gamma^y i \partial_x) \psi$ naively breaks translation and rotation symmetry according the symmetry transformations in Table~\ref{tab:symmetryTrans}. However, the fermion bilinear is invariant under a combination of microscopic symmetries and $\SU(2)$ gauge transformations:
\begin{equation}
  \begin{aligned}
    \mathscr{P} G \, : \, \overline{\psi}\sigma^a M\psi \rightarrow  \overline{\psi} U_G^\dagger V_G^\dagger \sigma^a V_G M U_G \psi \,.
  \end{aligned}
\end{equation}
$U_G$ is the action of the microscopic symmetry and $V_G$ is a gauge transformation, given by 
\begin{equation}
\begin{aligned}
  V_{tx}&=g(\phi_x)i\sigma^x,
  &
  V_{px}&=g(\phi_{px}),
  &
  V_{r}&=g(\phi_r)i\sigma^x,
  \\
  V_{ty}&=g(\phi_y)i\sigma^x,
  &
  V_{py}&=g(\phi_{py}),
  &
  V_t&=g(\phi_t),
  \label{eq:staggeredFluxProjective}
\end{aligned}
\end{equation}
where $g(\phi) \equiv e^{i \phi \sigma_z}$ reflects the residual $\UU(1)$ symmetry. The convention that we use for these subscripts are as follows. Translations along one lattice site in the $x$, $y$ direction are represented by subscripts $tx$, $ty$, respectively. Reflections along the $x$, $y$ axis are indicated by $px$, $py$. The subscript $r$ corresponds to a $\pi/2$ rotation, and $t$ indicates time-reversal symmetry. These projective symmetry transformations can be deduced by the way the microscopic symmetries act on the Dirac fermions, given in Table~\ref{tab:symmetryTrans}. The ability to condense additional Higgs fields to yield a $\mathbb{Z}_2$ spin liquid is constrained by the requirement that there exists a choice of phases $\phi_i$ such that all bilinear terms are invariant under the same projective symmetry transformation. We use this fact to identify the Yukawa couplings that correspond to the different possible proximate spin liquids - the phases $\phi_i$ can be read off via the symmetry fractionalization of the spin liquids, as shown in the next section, which uniquely identifies the Yukawa couplings consistent with these phases.

\subsection{Translationally-invariant spin liquid ansatzes}
There are four proximate $\mathbb{Z}_2$ spin liquids with translationally-invariant mean-field ansatzes - i.e., there exists a gauge in which translational symmetry is realized explicitly in the mean-field ansatz. These spin liquids are Z2A$zz$13, Z2A00$1n$, Z2A$zz1n$, and Z2A0013. These spin liquids are classified based off of their symmetry fractionalization; in other words, how symmetry operations like $T_y^{-1} T_x T_y T_x^{-1}$ are not directly equivalent to the identity, but equivalent up to a gauge transformation. These values are shown in Table~\ref{tab:symmetryFractionalization}, along with the values corresponding to the continuum $\UU(1)$ staggered flux phase, which can be read off from the transformations~\ref{eq:staggeredFluxProjective}.
\begin{table}[ht]
  \centering
  \caption{Symmetry fractionalization of the translationally-invariant $\mathbb{Z}_2$ spin liquids proximate to the staggered flux phase.}
  \label{tab:symmetryFractionalization}
  \begin{tabular}{ccccccc}
    \hline\hline
    & Group Relation & Staggered Flux & Z2A$zz$13 & Z2A0013 & Z2A$zz1n$ & Z2A00$1n$ \\
    \hline
    1 & $T_y^{-1} T_x T_y T_x^{-1}$ & $-e^{-2i(\phi_x - \phi_y) \sigma^z}$ & 1 & 1 & 1 & 1 \\
    2 & $P_y^{-1} T_x P_y T_x^{-1}$ & $e^{2 i \phi_{py} \sigma^z}$ & -1 & 1 & -1 & 1 \\
    3 & $P_y^{-1} T_y P_y T_y$ & $e^{2 i \phi_{py} \sigma^z}$ & -1 & 1 & -1 & 1 \\
    4 & $P_y^{2}$ & $e^{2 i \phi_{py} \sigma^z}$ & -1 & 1 & -1 & 1 \\
    5 & $P_y^{-1} R_{\pi / 2} P_y R_{\pi / 2}^{-1}$ & $-e^{2 i \phi_{py} \sigma^z}$ & 1 & -1 & 1 & -1 \\
    6 & $R_{\pi / 2}^4$ & $1$ & 1 & 1 & 1 & 1 \\
    7 & $R_{\pi /2}^{-1} T_x R_{\pi / 2} T_y $ & $e^{i(2 \phi_r - \phi_x - \phi_y)\sigma^z}$ & -1 & 1 & -1 & 1 \\
    8 & $R_{\pi /2}^{-1} T_y R_{\pi / 2} T_x^{-1} $ & $e^{i(2 \phi_r - \phi_x - \phi_y)\sigma^z}$ & -1 & 1 & -1 & 1 \\
    9 & $\mathcal{T}^{-1} R_{\pi /2}^{-1} \mathcal{T} R_{\pi /2}$ & $e^{-2i\phi_t \sigma^z}$ & -1 & -1 & 1 & 1 \\
    10 & $\mathcal{T}^{-1} P_{y}^{-1} \mathcal{T} P_{y}$ & 1 & 1 & 1 & 1 & 1 \\
    11 & $\mathcal{T}^{-1} T_{x}^{-1} \mathcal{T} T_{x}$ & $-e^{-2i\phi_t \sigma^z}$ & 1 & 1 & -1 & -1 \\
    12 & $\mathcal{T}^{-1} T_{y}^{-1} \mathcal{T} T_{y}$ & $-e^{-2i\phi_t \sigma^z}$ & 1 & 1 & -1 & -1 \\
    13 & $\mathcal{T}^2$ & $e^{-2i\phi_t \sigma^z}$ & -1 & -1 & 1 & 1 \\
  \end{tabular}
\end{table}
Using this table to identify the phases $\phi_i$ we find the continuum projective symmetry group for Z2A$zz$13,
\begin{equation}
  \begin{aligned}
    V_{tx} &= - i \sigma^y \,, \quad V_{px} = \pm i \sigma^z
    \\
    V_r &= - \frac{i}{\sqrt{2}} (\sigma^x - \sigma^y)
    \\
    V_{ty} &= - i \sigma^x \,, \quad V_{py} = - i \sigma^z \,, \quad V_t = i \sigma^z
  \end{aligned}
\end{equation}
Z2Azz1n,
\begin{equation}
  \begin{aligned}
    V_{tx} &= - i \sigma^y \,, \quad V_{px} = \pm i \sigma^z
    \\
    V_r &= - \frac{i}{\sqrt{2}} (\sigma^x - \sigma^y)
    \\
    V_{ty} &= - i \sigma^x \,, \quad V_{py} = - i \sigma^z \,, \quad V_t = 1
  \end{aligned}
\end{equation}
Z2A0013,
\begin{equation}
  \begin{aligned}
    V_{tx} &= - i \sigma^y \,, \quad V_{px} = 1
    \\
    V_r &= - \frac{i}{\sqrt{2}} (\sigma^x + \sigma^y)
    \\
    V_{ty} &= - i \sigma^x \,, \quad V_{py} = 1 \,, \quad V_t = i \sigma^z
  \end{aligned}
\end{equation}
and Z2A001n,
\begin{equation}
  \begin{aligned}
    V_{tx} &= - i \sigma^y \,, \quad V_{px} = 1
    \\
    V_r &= - \frac{i}{\sqrt{2}} (\sigma^x + \sigma^y)
    \\
    V_{ty} &= - i \sigma^x \,, \quad V_{py} = 1 \,, \quad V_t = 1\,.
  \end{aligned}
\end{equation}
From this, we can identify the Yukawa couplings consistent with these transformations, given in Table~\ref{tab:yukawaCouplingsZ2A}. In order to realize these transitions, we require \textit{two} Higgs fields, $\Phi_1$ and $\Phi_2$ which transform into each other under square lattice rotations and condense to acquire a non-zero $\langle\Phi_1^x\rangle$ and $\langle \Phi_2^y \rangle$. The fact that these fields are related under the microscopic rotational symmetry requires both Higgs fields to have the same mass, and hence they can both be condensed simultaneously by tuning a single parameter. These theories may be studied in a manner analogous to our study of the Z2A$zz$13 transition.
\begin{table}[ht]
  \centering
  \caption{Yukawa couplings of the two adjoint Higgs fields which realize the symmetry fractionalization of the proximate translationally-invariant spin liquids.}
  \label{tab:yukawaCouplingsZ2A}
  \begin{tabular}{ccccc}
    \hline\hline
  & Z2A$zz$13 & Z2A0013 & Z2A$zz1n$ & Z2A00$1n$ \\
  \hline
    $\Phi_1$ coupling & $\mu^x \gamma^y$ & $\mu^x \gamma^x$ & $\mu^x i \partial_x$ & $\mu^z i \partial_x$ \\
    $\Phi_2$ coupling & $\mu^z \gamma^x$ & $\mu^z \gamma^y$ & $\mu^z i \partial_y$ & $\mu^x i \partial_y$
  \end{tabular}
\end{table}
\subsection{Non-translationally-invariant spin liquid ansatzes}
The four additional spin liquid phases proximate to the staggered flux phase are Z2Bzz13, Z2B0013, Z2B001n, and Z2Bzz1n. These spin are distinct from the first four as one cannot write down a translationally-invariant mean-field ansatz for them - note that this does not correspond to a physical breaking of the translational symmetry, as the symmetry is still realized projectively; rather, the statement is that the symmetry \textit{must} be realized projectively. The symmetry fractionalization of these spin liquids is identical to their Z2A counterparts in Table~\ref{tab:symmetryFractionalization} aside from a change in sign in rows 1, 7, and 8. This leads to the continuum projective symmetry group for Z2Bzz13,
\begin{equation}
  \begin{aligned}
    V_{tx} &=  i \sigma^x \,, \quad V_{px} = \pm i \sigma^z
    \\
    V_r &=  i \sigma^x
    \\
    V_{ty} &=  i \sigma^x \,, \quad V_{py} = - i \sigma^z \,, \quad V_t = i \sigma^z
  \end{aligned}
\end{equation}
Z2Bzz1n,
\begin{equation}
  \begin{aligned}
    V_{tx} &=  i \sigma^x \,, \quad V_{px} = \pm i \sigma^z
    \\
    V_r &= i \sigma^x
    \\
    V_{ty} &= i \sigma^x \,, \quad V_{py} = - i \sigma^z \,, \quad V_t = 1
  \end{aligned}
\end{equation}
Z2B0013,
\begin{equation}
  \begin{aligned}
    V_{tx} &=  i \sigma^x \,, \quad V_{px} = 1
    \\
    V_r &= i \sigma^y
    \\
    V_{ty} &=  i \sigma^x \,, \quad V_{py} = 1 \,, \quad V_t = i \sigma^z
  \end{aligned}
\end{equation}
and Z2B001n,
\begin{equation}
  \begin{aligned}
    V_{tx} &=  i \sigma^x \,, \quad V_{px} = 1
    \\
    V_r &= i \sigma^y
    \\
    V_{ty} &=  i \sigma^x \,, \quad V_{py} = 1 \,, \quad V_t = 1\,.
  \end{aligned}
\end{equation}

Describing these phases as a Higgsed phase of the $\pi$-flux spin liquid turns out to differ significantly from their Z2A counterparts. The transition from the staggered flux phase to these spin liquids is driven by the condensation of a \textit{single} Higgs field $\Phi_1$, which acquires a non-zero expectation value $\langle \Phi_1^x \rangle$ in the $\mathbb{Z}_2$ phase. This prevents a direct transition from the $\SU(2)$ $\pi$-flux state to the $\mathbb{Z}_2$ spin liquid, as the masses of two Higgs fields $\Phi_1$ and $\Phi_3$ aren't constrained to be equal by the microscopic symmetries.

\begin{table}[ht]
  \centering
  \caption{Yukawa couplings of the single adjoint Higgs fields which realize the symmetry fractionalization of the proximate Z2B spin liquids.}
  \label{tab:yukawaCouplingsZ2B}
  \begin{tabular}{ccccc}
    \hline\hline
  & Z2Bzz13 & Z2B0013 & Z2Bzz1n & Z2B001n \\
  \hline
    $\Phi_1$ coupling & $\mu^y \partial_x \partial_y (\partial_x^2 + \partial_y^2)$ & $\mu^y$ & $i \partial_0$ & $\mu^y (\gamma^y i \partial_x - \gamma^x i \partial_y)$
  \end{tabular}
\end{table}

\section{Emergent subsystem symmetries in the \texorpdfstring{$\SU(2) \rightarrow \mathbb{Z}_2$}{SU(2) -> Z2} transition}
\label{app:subsystemSymmetry}
In this appendix, we provide a more detailed discussion of the emergent subsystem symmetry in the $\SU(2) \rightarrow \mathbb{Z}_2$ transition present in our theory. This transition is driven by the simultaneous condensation of the two Higgs fields $\Phi_{1, 2}$, whose masses are fixed to be equal by the microscopic $C_4$ rotation symmetry of the square lattice. The Lagrangian that describes this transition is
\begin{equation}
  \begin{aligned}
    \mathcal{L} &= \mathcal{L}_{\psi} + \mathcal{L}_{\Phi} + \mathcal{L}_{\Phi \psi}
    \\
    \mathcal{L}_{\psi} &= i \bar{\psi} \gamma^\mu \left( \partial_\mu - i A^a_\mu \sigma^a \right) \psi\,.
    \\
    \mathcal{L}_{\Phi} &= \sum_{i=1}^3 D_\mu \Phi_i^a D^\mu \Phi_i^a + V(\Phi)      
    \\
    \mathcal{L}_{\Phi \psi} &=  \Phi_1^a \, \bar{\psi} \mu^z \gamma^x \sigma^a \psi + \Phi_2^a \, \bar{\psi} \mu^x \gamma^y \sigma^a \psi
    \label{eq:phi12Lagrangian}
  \end{aligned}
\end{equation}
$V(\Phi)$ contains various symmetry-allowed potential terms for $\Phi_{1, 2}$, all of which are irrelevant other than the mass term which we tune to zero at criticality. Examining this action, we see that the quadratic fermion action along with the $\Phi_1$ Yukawa coupling in invariant under the transformation
\begin{equation}
\begin{aligned}
    \psi &\rightarrow e^{i  f_a(x) \mu^z \sigma^a}\psi 
    \\
    \Phi_1^a &\rightarrow U^{-1}_{ab}(x) \Phi_1^b + \partial_x f_a(x)
    \label{eq:subsystem1}
\end{aligned}
\end{equation}
and likewise for $\Phi_2$,
\begin{equation}
\begin{aligned}
    \psi &\rightarrow  e^{i  g_a(y) \mu^z \sigma^a} \psi
    \\
    \Phi_1^a &\rightarrow \tilde{U}^{-1}_{ab}(y) \Phi_1^b + \partial_y g_a(y)
    \label{eq:subsystem2}
    \end{aligned}
\end{equation}
with $U(x)$ the $\SO(3)$ rotation corresponding to the adjoint $\SU(2)$ action of $e^{i f_a(x) \sigma^a}$, i.e.
\begin{equation}
    U(x)_{ab} \equiv \frac{1}{2} \Tr \left[ e^{-i f_c(x) \sigma^c} \sigma^a e^{i f_d(x) \sigma^d} \sigma^b \right]
\end{equation}
and likewise for $\tilde{U}(y)$. The simplest way to see the existence of these symmetries is to consider $\Phi_1$ and $\Phi_2$ as the $x$ and $y$ components, respectively, of fictitious gauge fields corresponding to the $\SU(2)$ symmetries $\psi \rightarrow e^{i \sigma^a \mu^{z, x}} \psi$. The Yukawa couplings present in our theory couple these Higgs fields to the $x$ and $y$ components of the respective conserved currents. As such, the $\Phi_1$ ($\Phi_2$) Yukawa coupling and the fermion quadratic term are invariant under $x$-dependent ($y$-dependent) $\SU(2)$ transformations in a manner analogous to gauge invariance. Of course, this symmetry is ultimately broken in the full Lagrangian, both by the actual gauge field $A_\mu$ as well as the Higgs potential $V(\Phi)$; however, in a $1/N_f$ expansion, the leading-order effective action for the Higgs fields does possess this symmetry. This effective action is more relevant at long distances than subleading corrections, such as the bare Higgs action $\Phi_{1, 2}^a \partial^2 \Phi_{1, 2}^a$, and hence our analysis suggests that these emergent subsystem symmetries control the behavior of physical observables at criticality. The $\order{N_f^{-1}}$ contributions to the N\'eel and VBS order parameters have logarithm squared divergences, which we conjecture should lead to correlations decaying as $ \frac{1}{r^\alpha}e^{-\beta \ln^2(r)}$, with $\beta= -\frac{6}{\pi^2 N_f}$ for the VBS correlator and $-\frac{12}{\pi^2 N_f}$ for the N\'eel correlator, and $\alpha$ being some \textit{non-universal} coefficient. Interestingly, we note that the logarithm squared divergences for correlations of the scalar spin chirality exactly cancel at $\order{N_f^{-1}}$. We expect that this is related to the invariance of the scalar spin chirality order parameter under the subsystem symmetries given by Eqs.~(\ref{eq:subsystem1}) and (\ref{eq:subsystem2}). Many open questions remain in regards to these emergent symmetries; in particular, the physical interpretation of them is not clear - while these symmetry transformations are not purely gauge transformations, they do contain an action on the gauge $\SU(2)$ space given by the $\sigma^a$ operators, and as such, can rotate gauge-invariant operators such as $\bar{\psi} \mu^i \psi$ into non-gauge-invariant ones.
\section{Calculation of effective bosonic propagators}
\label{app:effectiveAction}
In order to study our critical theory in a ${1}/{N_f}$ expansion, we must calculate the effective propagators for both the Higgs and gauge bosons, which are generated by one-loop fermion diagrams. These calculations are complicated substantially by the non-Lorentz-invariant nature of the fermion propagator. As given in Eq.~(\ref{eq:sfLagrangian}), the bare fermion
action is given by
\begin{equation}
    \bar{\psi} \left[\slashed{k} + \Phi \mu^y \sigma^z (\gamma^y k_x + \gamma^x k_y) \right]\psi\,.
\end{equation}
It is useful to work in an eigenbasis of $\sigma^z \mu^y$. Inverting this, we get a diagonal $4 \times 4$ matrix in gauge/valley space, with elements
\begin{equation}
  \begin{aligned}
    \frac{\gamma^0 k_0  +\gamma^x (k_x \pm \Phi k_y) + \gamma^y (k_y \pm \Phi k_x) }{k_0^2 + (k_x \pm \Phi k_y)^2 + (k_y \pm \Phi k_x)^2} \,.
  \end{aligned}
\end{equation}
As in the main text, we define the variables
\begin{equation}
  \begin{aligned}
    k_{x, \pm} \equiv k_x \pm \Phi k_y \,, \quad k_{y, \pm} \equiv k_y \pm \Phi k_x\,, \abs{k_{\pm}} \equiv \sqrt{k_0^2 + k_{x, \pm}^2 + k_{y, \pm}^2}\,.
  \end{aligned}
\end{equation}
The inverse Higgs propagator is generated by the one-loop diagram given by Fig.~\ref{fig:effectiveActionDiagrams} of the main text, which translates to the integral
\begin{equation}
  \begin{aligned}
    &-\frac{4}{N_f} \sum_{a=\pm} \int \frac{\dd[3]{p}}{(2\pi)^3}  \frac{p_0(p_0 + k_0) +a p_{x, a} (p_{y, a} + k_{y, a}) + a p_{y, a}  (p_{x, a} + k_{x, a}))}{k_{a}^2 (k+p)_{a}^2 }
    \label{eq:higgsPropIntegrand2}
  \end{aligned}
\end{equation}
To simplify this integral, we perform a change of integration variables to $(p_0, p_{x, \pm}, p_{y, \pm})$. Performing this change of variables gives a factor of $(1-\Phi^2)^{-1}$ in the integral.

To evaluate this integral, we calculate the two general forms of integrals relevant to Eq.~(\ref{eq:higgsPropIntegrand2}).
\begin{equation}
  \begin{aligned}
    &\int \frac{\dd[3]{p}}{(2\pi)^3}\frac{p_0 (p_0 + k_0)}{p^2(p+k)^2}
      \\
      &= \int \frac{\dd[3]{p}}{(2\pi)^3}\frac{p_0(p_0 + k_0)}{\left[ (p + x k)^2 + \Delta \right]^2} \quad \Delta \equiv k^2 x(1-x)
      \\
      &= \int_0^1 \dd{x} \int \frac{\dd[3]{p}}{(2\pi)^3}\frac{(p_0 - x k_0) (p_0 + (1-x) k_0) }{(p^2 + \Delta)^2}
      \\
      &= \frac{1}{2\pi^2} \int_0^1 \dd{x} \int_0^\infty \dd{p} \frac{p^2 [p^2/3 - x(1-x) k_0^2]}{(p^2 + \Delta)^2}
      \\
      &= \frac{1}{2\pi^2} \int_0^1 \dd{x} \left[ -\frac{\pi \sqrt{\Delta}}{4} - x(1-x) \frac{k_0^2 \pi}{4 \sqrt{\Delta}} \right]
      \\
      &= - \frac{k^2 + k_0^2}{64 \abs{k}}
      \\
      &\int \frac{\dd[3]{p}}{(2\pi)^3}\frac{p_x (p_y + k_y)}{p^2(p+k)^2}
      \\
      &= \int_0^1 \dd{x} \int \frac{\dd[3]{p}}{(2\pi)^3}\frac{(p_x - x k_x) (p_y + (1-x) k_y) }{(p^2 + \Delta)^2}
      \\
    &= -\int_0^1 \dd{x} \int \frac{\dd[3]{p}}{(2\pi)^3}\frac{  x(1-x) k_x k_y }{(p^2 + \Delta)^2}
    \\
    &=- \frac{k_x k_y}{2\pi^2} \int_0^1 \dd{x} \int_0^\infty \dd{p} \frac{x (1-x)}{(p^2 + \Delta)^2}
    \\
    &= - \frac{k_x k_y}{64 \abs{k}}
  \end{aligned}
\end{equation}
In these calculations, we omit divergent terms which renormalize the boson mass, as this term is set to zero at criticality. Substituting this expression back into Eq.~(\ref{eq:higgsPropIntegrand}) for general $p_i, k_i$, we get the Higgs propagator
\begin{equation}
  \begin{aligned}
    \Gamma &= \frac{\lambda^2}{16 N_f (1-\Phi^2) } \left[ \frac{ k_+^2 + k_0^2 + 2 k_{x, +} k_{y, +}}{\abs{k_+}} + \frac{ k_-^2 + k_0^2 - 2 k_{x, -} k_{y, -}}{\abs{k_-}}     \right]
  \end{aligned}
\end{equation}

\section{Derivation of one-loop renormalization group equations}
\label{app:rg}
In this appendix, we give a derivation of the renormalization group equations used in the main text. The one-loop contributions to the fermion self-energy $\Sigma(k)$ are UV divergent, and hence require a UV cutoff $\Lambda$. The behavior of the self-energy upon integrating out high-energy modes is dictated by the logarithmic derivative with respect to the cutoff, $\Lambda \dv{\Lambda} \Sigma(k)$. The fact that our propagators are homogeneous functions of the three-momenta allow us to calculate this logarithmic derivative explicitly without reference to a specific cutoff. We assume that our regularized one-loop expression for the self-energy takes the form
\begin{equation}
  \begin{aligned}
    \Sigma(k) = \int \frac{\dd[3]{p}}{(2\pi)^3} F(p+k) G(p) \mathcal{K}\left( \frac{p^2}{\Lambda^2} \right) \mathcal{K}\left( \frac{(k+p)^2}{\Lambda^2} \right)
  \end{aligned}
\end{equation}
where $F$ and $G$ are homogeneous functions of the three-momenta with degree $-1$ - we take $F$ to be the fermion propagator, and $G$ to be the boson propagator (either Higgs or gauge) along with the various vertex coefficients. The function $\mathcal{K}(y)$ serves as a UV cutoff with the property that $\mathcal{K}(0) =1$ and $\mathcal{K}(y)$ falls off rapidly for large $y$, i.e., $\mathcal{K}(y) = e^{-y}$. Since we are interested in the behavior at small momenta, we expand around $k=0$, 
\begin{equation}
  \begin{aligned}
    \Sigma(k) \approx k_\mu \int \frac{\dd[3]{p}}{(2\pi)^3} \left[ \pdv{F(p)}{p_\mu} G(p) \mathcal{K}^2\left( \frac{p^2}{\Lambda^2} \right) + F(p) G(p) \mathcal{K}\left( \frac{p^2}{\Lambda^2} \right) \frac{2 p_\mu}{\Lambda^2} \mathcal{K}'\left(\frac{p^2}{\Lambda^2} \right) \right]\,.
  \end{aligned}
\end{equation}
We then take the logarithmic derivative,
\begin{equation}
  \begin{aligned}
  \Lambda \dv{\Lambda}  \Sigma(k) &\approx k_\mu \int \frac{\dd[3]{p}}{(2\pi)^3} \Bigg[ \left\{ - \frac{4 p^2}{\Lambda^2} \pdv{F(p)}{p_\mu} - 4 F(p) \frac{p_\mu}{\Lambda^2}\right\} G(p) \mathcal{K}\left( \frac{p^2}{\Lambda^2} \right)
  \mathcal{K}'\left( \frac{p^2}{\Lambda^2} \right)
  \\
  &- 4 \frac{p^2 p_\mu}{\Lambda^4} F(p) G(p) \left\{ \mathcal{K}\left( \frac{p^2}{\Lambda^2} \right)\mathcal{K}''\left( \frac{p^2}{\Lambda^2} \right)+ \mathcal{K}'^2\left( \frac{p^2}{\Lambda^2} \right) \right\}
  \Bigg]\,.
  \end{aligned}
\end{equation}
We now convert to spherical coordinates, $\vec{p} = y \Lambda ( \cos\theta, \sin\theta \sin\phi, \sin\theta \cos\phi)$, and use the homogeneity property of $F$ and $G$ to pull out factors of $(y \Lambda)^{-1}$. 
\begin{eqnarray}
\Lambda \frac{d}{d\Lambda} \Sigma (k) &\approx& \frac{ k_\mu}{8 \pi^3} \int_{0}^\pi \sin\theta \dd{\theta} \int_0^{2 \pi} \dd{\phi} \left[ \left\{- 4
 \frac{\partial F (\hat{p}) }{\partial p_\mu} - 4 \hat{p}_\mu F(\hat{p}) \right\} G (\hat{p}) \int_0^\infty y \dd{y} \mathcal{K}(y^2) \mathcal{K}' (y^2)   \right. \nonumber \\
 &~& \left.~~- 4 \hat{p}_\mu  F(\hat{p}) G (\hat{p}) 
 \int_0^\infty y^3 dy 
 \left\{
 \mathcal{K} \left(y^2 \right)  \mathcal{K}'' \left(y^2 \right) + \mathcal{K}^{\prime 2} \left(y^2 \right)\right\} \right]
\end{eqnarray}
The integral over $y$ can be done explicitly via integration by parts, which causes the dependence on the cutoff function $\mathcal{K}$ to drop out. This leads to the expression cited in the main text
\begin{equation}
  \begin{aligned}
    \Lambda \dv{\Lambda} \Sigma(k) &= \frac{k_\lambda}{8\pi^3}  \int_0^{2\pi} \dd{\phi} \int_0^\pi \sin\theta \dd{\theta} \pdv{F(\hat{p})}{p_\lambda} G(\hat{p})\,.
  \end{aligned}
\end{equation}
where $\hat{p} \equiv (\cos\theta, \sin\theta\sin\phi, \sin\theta \cos\phi)$.

Explicitly, we take, for the Higgs contribution to the self-energy, defining $Q(p)$ as the fermion propagator,
\begin{equation}
\begin{aligned}
    F(p) &=  \left(\mu^z \gamma^x + i \mu^x \gamma^y \right) \sigma^- Q(p) \left(\mu^z \gamma^x - i \mu^x \gamma^y \right) \sigma^+ + \left(\mu^z \gamma^x - i \mu^x \gamma^y \right) \sigma^+ Q(p) \left(\mu^z \gamma^x + i \mu^x \gamma^y \right) \sigma^-\,,
    \\
    G(p) &= \frac{1}{\Gamma(p)}\,.
    \end{aligned}
\end{equation}
For the gauge contribution,
\begin{equation}
\begin{aligned}
    F_{\mu\nu}(p) &= \left[ \gamma^\mu + \Phi \mu^y \sigma^z \left( \delta_{\mu, x} \gamma^y + \delta_{\mu, y} \gamma^x \right) \right] Q(p)\left[ \gamma^\nu + \Phi \mu^y \sigma^z \left( \delta_{\nu, x}  \gamma^y + \delta_{\nu, y}  \gamma^x \right) \right]\,,
    \\
    G^{\mu\nu}(p) &= \left(\Pi^{-1}(p)\right)^{\mu\nu}\,,
    \end{aligned}
\end{equation}
where the fermion self-energy now contains a summation over $\mu\,, \nu$.
A similar approach can be used to regulate the one-loop vertex corrections, which take the form
\begin{equation}
  \begin{aligned}
   \Xi_i  = \int \frac{\dd[3]{p}}{(2\pi)^3} H_i(p) \mathcal{K}^3\left( \frac{p^2}{\Lambda^2} \right)
  \end{aligned}
\end{equation}
where $H_i(p)$ is a homogeneous function of degree $-3$. 
Upon taking the logarithmic derivative,
\begin{equation}
  \begin{aligned}
   \Lambda \dv{\Lambda} \Xi_i  = -3 \int \frac{\dd[3]{p}}{(2\pi)^3} H_i(p) p^2 \mathcal{K}^2\left( \frac{p^2}{\Lambda^2} \right) \mathcal{K}'\left( \frac{p^2}{\Lambda^2} \right)
  \end{aligned}
\end{equation}

Converting to spherical coordinates and integrating by parts, we get
\begin{equation}
\Lambda \dv{\Lambda} \Xi_i = \frac{1}{8\pi^3}  \int_0^{2\pi} \dd{\phi} \int_0^\pi \sin\theta \dd{\theta} H_i(\hat{p}) \,.
\end{equation}
Explicitly, the Higgs correction is
\begin{equation}
    \begin{aligned}
        H_i(p) &= Q(p) \mu^i Q(p) \left(\mu^z \gamma^x + i \mu^x \gamma^y \right) \sigma^- \frac{1}{\Gamma(p)}\left(\mu^z \gamma^x - i \mu^x \gamma^y \right) \sigma^+
        \\
        &+ Q(p) \mu^i Q(p) \left(\mu^z \gamma^x -i \mu^x \gamma^y \right) \sigma^+ \frac{1}{\Gamma(p)}\left(\mu^z \gamma^x + i\mu^x \gamma^y \right) \sigma^-
    \end{aligned}
\end{equation}
and the gauge correction,
\begin{equation}
    \begin{aligned}
        H_i(p) &= Q(p) \mu^i Q(p)  \left[ \gamma^\mu + \Phi \mu^y \sigma^z \left( \delta_{\mu, x} \gamma^y + \delta_{\mu, y} \gamma^x \right) \right] \left(\Pi(p)^{-1}\right)^{\mu\nu}\left[ \gamma^\nu + \Phi \mu^y \sigma^z \left( \delta_{\nu, x} \gamma^y + \delta_{\nu, y} \gamma^x \right) \right]
    \end{aligned}
\end{equation}
\section{Anisotropic correlation functions in real space}
\label{app:ft}
As shown in the main text, the momentum-space N\'eel correlator is given by
\begin{equation}
  \begin{aligned}
    G_{\text{N\'eel}}(k) = \frac{1}{8 (1-\Phi^2)}\left( \abs{k_+} + \abs{k_{-}} \right) 
  \end{aligned}
\end{equation}
where we define $\abs{k_{\pm}} = \sqrt{k_0^2 + (k_x \pm \Phi k_y)^2 + (k_y \pm \Phi k_x)^2} $. The Fourier transform
of this function can be computed with the knowledge of the (suitably regularized) Fourier transform in three dimensions, $\abs{k} \rightarrow \frac{1}{r^4}$. We take the Fourier transform
\begin{equation}
  \begin{aligned}
    \int \frac{\dd[3]{k}}{(2\pi)^3} e^{i k \cdot r} \abs{k_\pm}
  \end{aligned}
\end{equation}
and perform a change of variables to shift the anisotropy to the spatial coordinates
\begin{equation}
  \begin{aligned}
    &\frac{1}{1-\Phi^2} \int \frac{\dd[3]{k}}{(2\pi)^3} e^{i k \cdot r_{\pm}} \abs{k}
    \\
    t_{\pm} &= t \\
    x_{\pm} &= \frac{1}{1-\Phi^2}(x \pm \Phi y) \\
    y_{\pm} &= \frac{1}{1-\Phi^2}(y \pm \Phi x) \\
  \end{aligned}
\end{equation}
which yields the real space correlator given in the main text.

To compute the Fourier transform of the VBS correlator, given perturbatively by
\begin{equation}
  \begin{aligned}
    G_{\text{VBS}}(k) \sim \abs{k} - \Phi^2 \left[ 2 \abs{k} + \frac{k_x^2 k_y^2 - k_0^2 k^2}{\abs{k}^3} \right] + \order{\Phi^4}\,,
  \end{aligned}
\end{equation}
we define the function
\begin{equation}
  \begin{aligned}
    f(a_i, k_i) = \sqrt{a_0 k_0^2 + a_x k_x^2 + a_y k_y^2} \,.
  \end{aligned}
\end{equation}
The Fourier transform of this function can be calculated by a similar change of variables,
\begin{equation}
  \begin{aligned}
    \int \frac{\dd[3]{k}}{(2\pi)^3} f(a_i, k_i) \sim \frac{1}{\sqrt{a_0 a_x a_y} }\left( \frac{t^2}{a_0} + \frac{x^2}{a_x} + \frac{y^2}{a_y} \right)^{-2}\,.
  \end{aligned}
\end{equation}
The various terms in the $\order{\Phi^2}$ corrections to the VBS correlator can be obtained by taking derivatives
of $f(a_i, k_i)$ with respect to $a_i$ and setting $a_i = 1$. This allows us to calculate the real space VBS correlator
and gives the result in the main text.

\section{Perturbative corrections to monopole scaling dimension}
\label{app:monopole}
In this appendix, we present a partial calculation of the $\order{N_f}$ corrections to the scaling dimensions of a monopole at our deconfined critical point. As previously established~\cite{Borokhov2002}, the $\order{N_f}$ scaling dimension for isotropic $\text{QED}_3$ is $\Delta = 1.06 N_f$. We present the scaling dimension using our convention, where $\text{QED}_3$ with $N=4$ Dirac fermions corresponds to $N_f = 1$. Although gauge and Higgs fluctuations give corrections to this value, these corrections are subleading in $N_f$, and the
only $\order{N_f}$ correction comes from taking the saddle-point solutions of the bosonic fields and calculating the shift in free energy arising from the anisotropic Dirac dispersion relation. We proceed perturbatively in the Dirac anisotropy parameter $\Phi$ - this is necessary as the anisotropy will in principle modify the saddle-point monopole configuration of the gauge field. As we will see, to leading order in $\Phi$, the gauge field configuration corresponding to the isotropic Dirac monopole will be sufficient.

We start with the action for $\text{QED}_3$ with the allowed velocity anisotropy term, omitting the Higgs fields as they will not play any role in the calculation
\begin{equation}
    S = i \int \dd[3]{r} \left[ \bar{\psi}  \slashed{D} \psi + \Phi \bar{\psi} \mu^y \left(\gamma^x D_x - \gamma^y D_y \right) \right] \psi\,.
    \label{eq:saddlePointAction}
\end{equation}
 We leave implicit the summation over the $4 N_f$ fermions. Note that this action is different than in the main text. This is because we follow the convention used in~\cite{Hermele05}, where the gauge field is coupled in the usual way, $D_\mu \equiv \partial_\mu - i A_\mu$, and the microscopic $\SU(2)$ spin rotation symmetry is implemented explicitly by the $\sigma^i$ matrices. We refrain from using this convention in the main calculation, as the coupling to the Higgs field is not easily expressible in this form and overall makes the calculation more complicated.

In the absence of a velocity anisotropy, the saddle-point configurations for the gauge field corresponding to $n$ units of magnetic flux at the origin are given by
\begin{equation}
    \bar{A}_n(r) = \frac{n}{2} (1-\cos \theta) \dd{\phi}
\end{equation}
The non-zero anisotropy will affect these saddle-point solutions. The leading order corrections to these solutions are $\order{\Phi^2}$, as the $\order{N_f}$ effective action for the gauge field upon integrating out the fermions has only corrections of $\order{\Phi^2}$ and higher. Hence, we write the saddle-point gauge field solution in the presence of a velocity anisotropy as $A_n(r) \equiv \bar{A}_n(r) + \delta A_n$, with $\delta A_n \sim \order{\Phi^2}$.

In order to calculate the scaling dimension of this monopole, we set $r = e^\tau$ and perform a Weyl rescaling
\begin{equation}
    \begin{aligned}
        g_{\mu \nu} &\rightarrow e^{-2\tau} g_{\mu \nu}
        \\
        \bar{\psi} \,, \psi &\rightarrow e^{-\tau} \bar{\psi}\,, e^{-\tau} \psi
    \end{aligned}
\end{equation}
This rescaling maps the scaling dimension of the monopole operator to the free energy $F = - \log Z$ of the system~\cite{Metlitski2008}. 

To leading order in $N_f$, we ignore gauge and Higgs fluctuations, and the action reduces down to one of free fermions with a background gauge field. This action can be put in a nearly-diagonal form with the aid of monopole harmonics~\cite{WU1976} and their spinor generalization~\cite{Borokhov2002}. By expanding $\psi$ in terms of these harmonics,
\begin{equation}
    \psi(r) = \int \frac{\dd{\omega}}{2\pi} \left[ \sum_{\ell = n/2}^\infty \sum_{m=-\ell-1}^\ell \Psi_T^{\ell m}(\omega) T_{n, \ell m}(\theta, \phi) + \sum_{\ell = n/2}^\infty \sum_{m=-\ell}^\ell \Psi_S^{\ell m}(\omega) S_{n, \ell m}(\theta, \phi) \right] e^{-i \omega \tau}
\end{equation}
where $T_{n, \ell m} \,, S_{n, \ell m}$ are eigenvalues of the orbital angular momentum operator $\vec{L}^2$ in the presence of a strength $n$ monopole, with orbital angular momentum $\ell$ and total angular momentum $\ell + 1/2$ for $T_{n, \ell m}$ and $\ell - 1/2$ for $S_{n, \ell m}$. Explicit expressions for the spinor harmonics $T_{n, \ell m}$ and $S_{n, \ell m}$ may be found in~\cite{Borokhov2002}. The variables $\Psi_T^{\ell m}\,, \Psi_S^{\ell m}$ are anti-commuting coefficients.

Expanded in this form, the isotropic action with $\Phi=0$ can be written as
\begin{equation}
    S_0 = \int \frac{\dd{\omega}}{2\pi} \sum_{\ell = n/2}^\infty \sum_{m=-\ell}^{\ell-1} \begin{pmatrix} \Psi_T^{(\ell-1) m} (\omega)^* &  \Psi_S^{\ell m}(\omega)^* \end{pmatrix} \vb{N}_{n, \ell} (\omega + i \vb{M}_{n, \ell})\begin{pmatrix} \Psi_T^{(\ell-1) m} (\omega) 
    \\ \Psi_S^{\ell m}(\omega) \end{pmatrix} 
\end{equation}
with
\begin{equation}
    \begin{aligned}
        \vb{M}_{n, \ell} &= \begin{pmatrix} \ell \left( 1 - \frac{n^2}{4 \ell^2} \right) & - \frac{n}{2} \sqrt{1 - \frac{n^2}{4 \ell^2}} \\ - \frac{n}{2} \sqrt{1 - \frac{n^2}{4 \ell^2}} & \ell \left( 1 - \frac{n^2}{4 \ell^2} \right) \end{pmatrix}\,,
        \\
        \vb{N}_{n, \ell} &= \begin{pmatrix} - \frac{n}{2\ell} & - \sqrt{1 - \frac{n^2}{4 \ell^2} } \\- \sqrt{1 - \frac{n^2}{4 \ell^2} } &   \frac{n}{2\ell} \end{pmatrix}\,.
    \end{aligned}
\end{equation}
The free energy is then given by
\begin{equation}
    \log Z_0 = 4 N_f \int \frac{\dd{\omega}}{2\pi} \sum_{\ell=n/2}^\infty \log \det (\vb{N}_{n, \ell} (\omega + i \vb{M}_{n, \ell} ))\,.
\end{equation}
This expression can be evaluated via zeta function regularization, yielding the aforementioned scaling dimension of $\Delta = 1.06 N_f$.

Corrections to the free energy as a consequence of a non-zero $\Phi$ can be calculated perturbatively. Writing the action as $S = S_0 + \delta S$, with $\delta S \sim \order{\Phi}$, we have 
\begin{equation}
\begin{aligned}
    \log Z &= \log \left[ \int \mathcal{D} \psi \mathcal{D} \bar{\psi} e^{-S} \right] = 
    \log \left[ Z_0 + \int \mathcal{D} \psi \mathcal{D} \bar{\psi} e^{-S_0} \left( - \delta S + \frac{1}{2} \delta S^2 + \ldots \right) \right]
    \\
    &= \log Z_0 - \langle \delta S \rangle + \frac{1}{2} \langle \delta S^2 \rangle + \ldots\,,
\end{aligned}  
\end{equation}
where the expectation values are evaluated with the isotropic action. There are two components of $\delta S$ that are $\order{\Phi^2}$ or lower, as can be seen from Eq.~(\ref{eq:saddlePointAction}). The first comes from the velocity anisotropy term $\Phi \bar{\psi} \mu^y (\gamma^x \bar{D}_x - \gamma^y \bar{D}_y ) \psi$, where the bar indicates that the covariant derivative is defined with the isotropic monopole gauge configuration. The second component arises from $\order{\Phi^2}$ corrections to the isotropic gauge configuration, which appear in the term $\bar{\psi} \slashed{D} \psi$. Only the first of these corrections gives $\order{\Phi^2}$ contributions to the free energy; the corrections to the saddle-point gauge configuration $\delta A_n$ couple to the conserved current $J^\mu = \bar{\psi} \gamma^\mu \psi + \order{\Phi}$, whose expectation value vanishes. Hence, the free energy to $\order{\Phi^2}$ is given by
\begin{equation}
\begin{aligned}
    F &= \log Z_0 + 2 \Phi^2\Tr \int \dd[3]{r} \dd[3]{r'} \langle \bar{\psi}(r) (\gamma^x (\bar{D}_x - \hat{x}) - \gamma^y (\bar{D}_y - \hat{y})) \psi(r)  
    \\
    & \times  \bar{\psi}(r') (\gamma^x (\bar{D}_x - \hat{x}) - \gamma^y (\bar{D}_y - \hat{y})) \psi(r') \rangle
    \label{eq:freeEnergyCorrection}
    \end{aligned}
\end{equation}
The additional factors of $\hat{x}\,, \hat{y}$ arise from the Weyl rescaling.

The calculation in Eq.~(\ref{eq:freeEnergyCorrection}) amounts to calculating the two-point function of the fermion bilinear $ \bar{\psi} (r) (\gamma^y (\bar{D}_x - \hat{x}) + \gamma^x (\bar{D}_y - \hat{y})) \psi(r)$. To this end, we denote the matrix elements of the operator $ (\gamma^y (\bar{D}_x - \hat{x}) + \gamma^x (\bar{D}_y - \hat{y}))$ in the spinor harmonic basis
\begin{equation}
   B_{n, \ell m \ell' m'}(\omega) = \int \dd{\Omega} \begin{pmatrix} T^\dagger_{n, \ell m} (\theta, \phi)e^{i \omega \tau}  \\ S^\dagger_{n, \ell m}(\theta, \phi) e^{i \omega \tau} \end{pmatrix} \left[\gamma^y (\bar{D}_x - \hat{x}) + \gamma^x (\bar{D}_y - \hat{y}) \right] \begin{pmatrix} T_{n, \ell m}(\theta, \phi)e^{-i \omega \tau}  & S_{n, \ell m}(\theta, \phi)e^{-i \omega \tau}  \end{pmatrix}
\end{equation}
These functions are exactly calculable in terms of Wigner $3$-$j$  symbols, and are only non-zero for $\abs{\ell - \ell'}\,, \abs{m - m'} \leq 2$. 
In order to calculate these functions, we need the matrix elements
\begin{equation}
  \begin{aligned}
    &\bra{Y_{q, \ell m}} \hat{x} \ket{Y_{q, \ell' m'}}
    \\
    &\bra{Y_{q, \ell m}} \hat{y} \ket{Y_{q, \ell' m'}}
    \\
    &\bra{Y_{q, \ell m}} D^\bot_x \ket{Y_{q, \ell' m'}}
    \\
    &\bra{Y_{q, \ell m}} D^\bot_y \ket{Y_{q, \ell' m'}}
  \end{aligned}
\end{equation}
where $Y_{q, \ell m}$ is the scalar monopole harmonic~\cite{WU1976} in a background monopole of strength $ 2 q \equiv n$ and $D^\bot_i$ is the angular component of the covariant derivative $D_i$; the radial component is simply equal to $\hat{i} \pdv{\tau}$.
For this, we need the integral formula for three monopole harmonics
\begin{equation}
  \begin{aligned}
    \int \dd{\hat{n}} Y_{q, \ell m} Y_{q', \ell' m'} Y_{q'', \ell'' m''} = (-1)^{\ell+\ell'+\ell''} \sqrt{\frac{(2\ell+1)(2\ell'+1)(2\ell''+1)}{4\pi}} \tj{\ell}{\ell'}{\ell''}{q}{q'}{q''} \tj{\ell}{\ell'}{\ell''}{m}{m'}{m''}
  \end{aligned}
\end{equation}
The first two matrix elements can be easily computed with the identity
\begin{equation}
  \begin{aligned}
    \hat{x} &= - \sqrt{\frac{4\pi}{6}} \left( Y_{0, l, 1} - Y_{0, 1, -1} \right)  \\
    \hat{y} &= i\sqrt{\frac{4\pi}{6}} \left( Y_{0, l, 1} + Y_{0, 1, -1} \right)  \\
    \bra{Y_{q, \ell, m}} \hat{x} \ket{Y_{q, \ell', m'}} &= (-1)^{\ell+\ell'+q+m}\sqrt{\frac{(2\ell+1)(2\ell'+1)}{2}}  \tj{\ell}{1}{\ell'}{-q}{0}{q} \left[ \tj{\ell}{1}{\ell'}{-m}{1}{m'}- \tj{\ell}{1}{\ell'}{-m}{-1}{m'} \right]  \\
    \bra{Y_{q, \ell, m}} \hat{y} \ket{Y_{q, \ell', m'}} &=-i (-1)^{\ell+\ell'+q+m}\sqrt{\frac{(2\ell+1)(2\ell'+1)}{2}}  \tj{\ell}{1}{\ell'}{-q}{0}{q} \left[ \tj{\ell}{1}{\ell'}{-m}{1}{m'}+ \tj{\ell}{1}{\ell'}{-m}{-1}{m'} \right]  \\
  \end{aligned}
\end{equation}
To calculate the last two matrix elements, we must utilize the raising and lowering angular momenta operators
\begin{equation}
  \begin{aligned}
    z L_x =z( L_+ + L_-) = D_x^\bot - \hat{y} \pdv{\phi}
  \end{aligned}
\end{equation}
and similarly for $D_y^\bot$.
This equation can be easily verified for the angular momenta operators without a monopole background, and we verify numerically that this formula correctly generalizes to non-zero $q$. This leads to the formula
\begin{equation}
  \begin{aligned}
    &\bra{Y_{q, \ell m}} D^\bot_x \ket{Y_{q, \ell' m'}} = (-1)^{1+\ell+\ell'+q+m} \sqrt{\frac{(2\ell+1)(2\ell'+1)}{4}} \tj{\ell}{1}{\ell'}{-q}{0}{q} 
    \\
    &\times\left[\sqrt{(\ell'-m')(\ell'+m'+1)} \tj{\ell}{1}{\ell'}{-m}{0}{m'+1} - \sqrt{(\ell'+m')(\ell'-m'+1)} \tj{\ell}{1}{\ell'}{-m}{0}{m'-1}\right]   \\
    - &i m' \bra{Y_{q, \ell, m}} \hat{y} \ket{Y_{q, \ell', m'}}
    \\
    &\bra{Y_{q, \ell m}} D^\bot_y \ket{Y_{q, \ell' m'}} = -i (-1)^{1+\ell+\ell'+q+m} \sqrt{\frac{(2\ell+1)(2\ell'+1)}{4}} \tj{\ell}{1}{\ell'}{-q}{0}{q} 
    \\
    &\times\left[\sqrt{(\ell'-m')(\ell'+m'+1)} \tj{\ell}{1}{\ell'}{-m}{0}{m'+1} + \sqrt{(\ell'+m')(\ell'-m'+1)} \tj{\ell}{1}{\ell'}{-m}{0}{m'-1}\right]   \\
    - &i m' \bra{Y_{q, \ell m}} \hat{y} \ket{Y_{q, \ell' m'}}
  \end{aligned}
\end{equation}
From these matrix elements, the components of $B_{n, \ell m \ell' m'}$ can be assembled by expressing the spinor monopole harmonics in terms of the scalar harmonics.

Upon obtaining an expression for $B$, we have
\begin{equation}
    \begin{aligned}
        F = \log Z_0 - 4\Phi^2 \sum_{\ell, \ell', m, m'} \int \frac{\dd{\omega}}{2\pi}\Tr \left[ B_{n, \ell m \ell' m'}(\omega) \vb{N}_{n, \ell}( \omega + i \vb{M}_{n, \ell} )B^\dagger_{n, \ell' m' \ell m}(\omega)\vb{N}_{n, \ell'}(\omega - i \vb{M}_{n, \ell'} ) \right]
        \label{eq:monopoleFreeEnergy}
    \end{aligned}
\end{equation}
The minus sign outside the summation relative to Eq.~\ref{eq:freeEnergyCorrection} arises from the fermion loop. What remains is a suitable procedure for regularizing the divergent expression in Eq.~\ref{eq:monopoleFreeEnergy} - as the functions $B_{n, \ell m \ell' m'}$ are rather complicated summations of Wigner $3$-$j$ symbols, this is a non-trivial task and we leave this as an open question for future study.

\bibliography{neelu1}

\end{document}